\def\etal{{\it et al.}}
\def\ie{{\it i.e.}}

\def\~{{$\tilde{\phantom{a}}$}}

\documentclass [12pt] {article}
\usepackage{epsfig}
\usepackage{color}

\def\thebibliography#1{\section{References}\markboth
 {REFERENCES}{REFERENCES}\list
 {[\arabic{enumi}]}{\settowidth\labelwidth{[#1]}\leftmargin\labelwidth
 \advance\leftmargin\labelsep
 \usecounter{enumi}}
 \def\newblock{\hskip .11em plus .33em minus -.07em}
 \sloppy
 \sfcode`\.=1000\relax}
\def\upcite#1{\raise6pt\hbox{\scriptsize
\cite{#1}}}
  \def\lsim{\mathrel {\vcenter {\baselineskip 0pt \kern 0pt
    \hbox{$<$} \kern 0pt \hbox{$\sim$} }}}
    \def\gsim{\mathrel {\vcenter {\baselineskip 0pt \kern 0pt
    \hbox{$>$} \kern 0pt \hbox{$\sim$} }}}

\def\hline{\noalign{\hrule \vskip2pt}}

\def\|{\ifmmode\Vert\else \char`\|\fi}
\ifx\oldzeta\undefined                          
  \let\oldzeta=\zeta                            
  \def\zzeta{{\raise 2pt\hbox{$\oldzeta$}}}     
  \let\zeta=\zzeta                              
\fi

\ifx\oldchi\undefined                           
  \let\oldchi=\chi                              
  \def\cchi{{\raise 2pt\hbox{$\oldchi$}}}       
  \let\chi=\cchi                                
\fi

\def\frac#1#2{{#1 \over #2}}

\def\half{\ifinner {\scriptstyle {1 \over 2}}
   \else {1 \over 2} \fi}

\def\abs#1{\left\vert#1\right\vert}	

\def\simge{\mathrel{%
   \rlap{\raise 0.511ex \hbox{$>$}}{\lower 0.511ex \hbox{$\sim$}}}}
\def\simle{\mathrel{
   \rlap{\raise 0.511ex \hbox{$<$}}{\lower 0.511ex \hbox{$\sim$}}}}

\def\buildchar#1#2#3{{\null\!                   
   \mathop#1\limits^{#2}_{#3}                   
   \!\null}}                                    
\def\overcirc#1{\buildchar{#1}{\circ}{}}

\def\slashchar#1{\setbox0=\hbox{$#1$}           
   \dimen0=\wd0                                 
   \setbox1=\hbox{/} \dimen1=\wd1               
   \ifdim\dimen0>\dimen1                        
      \rlap{\hbox to \dimen0{\hfil/\hfil}}      
      #1                                        
   \else                                        
      \rlap{\hbox to \dimen1{\hfil$#1$\hfil}}   
      /                                         
   \fi}                                         %

\def\subrightarrow#1{
  \setbox0=\hbox{
    $\displaystyle\mathop{}
    \limits_{#1}$}
  \dimen0=\wd0
  \advance \dimen0 by .5em
  \mathrel{
    \mathop{\hbox to \dimen0{\rightarrowfill}}
       \limits_{#1}}}                           

\def\overlay#1#2{\ifmmode%
\setbox0=\hbox{$#1$}%
\setbox1=\hbox to\wd0{\hss$#2$\hss}\else%
\setbox0=\hbox{#1}%
\setbox1=\hbox to\wd0{\hss#2\hss}\fi%
#1\hskip-\wd0\box1 }

\def\pmb#1{\leavevmode\setbox0=\hbox{#1}%
\kern-.02em\copy0\kern-\wd0
\kern.04em\copy0\kern-\wd0
\kern-.02em\raise.04em\box0 }

\def\vereq#1#2{\lower3pt\vbox{\baselineskip1.5pt \lineskip1.5pt
\ialign{$\m@th#1\hfill##\hfil$\crcr#2\crcr\sim\crcr}}}

\def\tensor#1{\protect\@ontopof{#1}{\leftrightarrow}{1.15}\mathord{\box2}}
\def\overstar#1{\protect\@ontopof{#1}{\ast}{1.15}\mathord{\box2}}
\def\overdots#1{\protect\@ontopof{#1}{\cdots}{1.0}\mathord{\box2}}
\def\overcirc#1{\protect\@ontopof{#1}{\circ}{1.2}\mathord{\box2}}
\def\loarrow#1{\protect\@ontopof{#1}{\leftarrow}{1.15}\mathord{\box2}}
\def\roarrow#1{\protect\@ontopof{#1}{\rightarrow}{1.15}\mathord{\box2}}

\def\@ontopof#1#2#3{%
{\mathchoice
{\@@ontopof{#1}{#2}{#3}\displaystyle\scriptstyle}%
{\@@ontopof{#1}{#2}{#3}\textstyle\scriptstyle}%
{\@@ontopof{#1}{#2}{#3}\scriptstyle\scriptscriptstyle}%
{\@@ontopof{#1}{#2}{#3}\scriptscriptstyle\scriptscriptstyle}%
}%
}

\def\@@ontopof#1#2#3#4#5{%
\setbox0=\hbox{$#4#1$}%
\setbox1=\hbox{$#5#2$}%
\setbox2=\hbox{}\ht2=\ht0 \dp2=\dp0 %
\ifdim\wd0>\wd1 %
\setbox1=\hbox to\wd0{\hss\box1\hss}%
\mathord{\rlap{\raise#3\ht0\box1}\box0}%
\else   %
\setbox1=\hbox to.9\wd1{\hss\box1\hss}%
\setbox0=\hbox to\wd1{\hss$#4\relax#1$\hss}%
\mathord{\rlap{\copy0}\raise#3\ht0\box1}%
\fi
}%

\def\lambdabar{\protect\@lambdabar}
\def\@lambdabar{%
\relax
\bgroup
\def\@tempa{\hbox{\raise.73\ht0
\hbox to0pt{\kern.25\wd0\vrule width.5\wd0
height.1pt depth.1pt\hss}\box0}}%
\mathchoice{\setbox0\hbox{$\displaystyle\lambda$}\@tempa}%
{\setbox0\hbox{$\textstyle\lambda$}\@tempa}%
{\setbox0\hbox{$\scriptstyle\lambda$}\@tempa}%
{\setbox0\hbox{$\scriptscriptstyle\lambda$}\@tempa}%
\egroup
}

\def\corresponds{{\lower.2ex\hbox{=}}{\rm\kern-.75em^\triangle}}
\def\succsim{\succ\kern-.9em_\sim\kern.3em}
\def\precsim{\prec\kern-1em_\sim\kern.3em}
\def\slantfrac#1#2{\kern1em^{#1}\kern-.3em/\kern-.1em_{#2}}

\begin{document}

\begin{center}
{\Large\bf Negative Group Velocity}
\\

\medskip

Kirk T.~McDonald
\\
{\sl Joseph Henry Laboratories, Princeton University, Princeton, NJ 08544}
\\
(July 23, 2000)
\end{center}

\section{Problem}

Consider a variant on the physical situation of 
``slow light'' \cite{Hau,slowlight} in which two closely spaced spectral
lines are now both optically pumped to show that the group velocity can be
negative at the central frequency, which leads to apparent superluminal
behavior.

\subsection{Negative Group Velocity}

In more detail, consider a classical model of matter in which spectral lines
are associated with oscillators.  In particular, consider a gas with two
closely spaced spectral lines of angular frequencies
 $\omega_{1,2} = \omega_0 \pm \Delta/2$, where
$\Delta \ll \omega_0$.  Each line has
the same damping constant (and spectral width) $\gamma$.  

Ordinarily, the gas would exhibit strong absorption of light in the
vicinity of the spectral lines.  But suppose that lasers of frequencies 
$\omega_1$ and $\omega_2$
pump the both oscillators into inverted populations.  This can be described 
classically by assigning negative oscillator strengths to
these oscillators \cite{note}.

Deduce an expression for the group velocity $v_g(\omega_0)$ of a pulse 
of light centered on frequency $\omega_0$ in terms of the (univalent)
plasma frequency $\omega_p$ of the medium, given by
\begin{equation}
\omega_p^2 = {4 \pi N e^2 \over m},
\label{eq0}
\end{equation}
where $N$ is the number density of atoms, and $e$ and $m$ are the charge and
mass of an electron.  Give a condition on the line
separation $\Delta$ compared to the line width $\gamma$ such that the
group velocity $v_g(\omega_0)$ is negative.

In a recent experiment by Wang \etal\ \cite{Wang}, a group velocity of 
$v_g = - c/310$, where $c$ is the speed of light in vacuum,
was demonstrated in cesium vapor using a pair of spectral lines with
separation $\Delta / 2 \pi \approx 2$ MHz and linewidth $\gamma / 2 \pi 
\approx 0.8$ MHz.

\subsection{Propagation of a Monochromatic Plane Wave}

Consider a wave with electric field $E_0 e^{i \omega(z / c - t)}$ that
is incident from $ z < 0$ on a medium that extends from 
$z = 0$ to $a$.  Ignore reflection at the boundaries, as is reasonable if
the index of refraction $n(\omega)$ is near unity.
Particularly simple results can be obtained when you make
the (unphysical) assumption that the $\omega n(\omega)$ varies linearly 
with frequency about a central frequency $\omega_0$.
Deduce a transformation that has a frequency-dependent part and a 
frequency-independent part between the phase of the
wave for $z < 0$ to that of the wave inside the medium, and to that of the
wave in the region $a < z$.    
 
\subsection{Fourier Analysis}

Apply the transformations between an incident monochromatic wave and
the wave in and beyond the medium to the Fourier analysis of an incident pulse
of form $f(z/c - t)$.

\subsection{Propagation of a Sharp Wave Front}

In the approximation that $\omega n$ varies linearly with $\omega$, deduce 
the waveforms in the regions $0 < z < a$ and $a < z$ for an incident
pulse $\delta(z/c - t)$, where $\delta$ is the Dirac delta function.
Show that the pulse emerges out of the gain region at $z = a$ at
time $t = a/v_g$, which appears to be earlier than when it enters this region
if the group velocity is negative.  Show also that inside the negative
group velocity medium a
pulse propagates backwards from $z = a$ at time $t = a/v_g < 0$ to $z = 0$
at $t = 0$, at which time it appears to annihilate the incident pulse.

\subsection{Propagation of a Gaussian Pulse}

As a more physical example, deduce the waveforms in the regions $0 < z < a$ 
and $a < z$ for a Gaussian incident pulse 
$E_0 e^{-(z/c - t)^2 / 2 \tau^2} e^{i \omega_0(z / c - t)}$.  Carry the
frequency expansion of $\omega n(\omega)$ to second order to obtain
conditions of validity of the analysis such as maximum pulsewidth $\tau$,
maximum length $a$ of the gain region, and maximum time of advance of
the emerging pulse.  Consider the time required to generate a pulse of
risetime $\tau$ when assessing whether the time advance in a negative group
velocity medium can lead to superluminal signal propagation.

\section{Solution}

The concept of group velocity appears to have been first enunciated by
Hamilton in 1839 in lectures of which only abstracts were published 
\cite{Hamilton}.  The first recorded observation of the group velocity of a 
(water) wave is due to Russell in 1844 \cite{Russell}.  However, widespread
awareness of the group velocity dates from 1876 when Stokes used its as the
topic of a Smith's Prize examination paper \cite{Stokes}.  The early history
of group velocity has been reviewed by Havelock \cite{Havelock}.

H.~Lamb \cite{Lamb} credits A.~Schuster with noting in 1904
 that a negative group
velocity, \ie, a group velocity of opposite sign to that of the phase velocity,
is possible due to anomalous dispersion.  Von Laue \cite{vonLaue}
made a similar comment in 1905.
Lamb gave two examples
of strings subject to external potentials that exhibit negative group
velocities.  These early considerations assumed that in case of a wave with
positive group and phase velocities incident on the anomalous medium, 
energy would be 
transported into the medium with a positive group velocity, and so there
would be waves with negative phase velocity inside the medium.  Such 
negative phase velocity waves are formally consistent with Snell's law
 \cite{Mandelstam} (since $\theta_t = \sin^{-1}[(n_i/n_t) \sin\theta_i]$ can be
in either the first or second quadrant), but they seemed physically 
implausible and the topic was largely dropped.

Present interest in negative group velocity a based on anomalous dispersion
in a gain medium, where the sign of the phase velocity is the same for
incident and transmitted waves, and energy flows inside the gain medium
in the opposite direction to the incident energy flow in vacuum. 

The propagation of electromagnetic waves at frequencies near those of
spectral lines of a medium was first extensively discussed by Sommerfeld and
Brillouin \cite{Brillouin}, with emphasis on the distinction between
signal velocity and group velocity when the latter exceeds $c$.
The solution presented here is based on the work of Garrett and
McCumber \cite{Garrett}, as extended by Chiao \etal\  \cite{Chiao93,Bolda94}.
A discussion of negative group velocity in electronic circuits has been given
by Mitchell and Chiao \cite{Mitchell}.

\subsection{Negative Group Velocity}

 In a medium of index of refraction $n(\omega)$, the 
dispersion relation can be written
\begin{equation}
k = {\omega n \over c},
\label{eq1}
\end{equation}
where $k$ is the wave number.  The group velocity is then given by
\begin{equation}
v_g = Re \left[ {d\omega \over dk} \right] 
= {1 \over Re[ dk/d\omega]} 
= {c \over Re[d (\omega n) / d\omega]}
= {c \over n + \omega Re[dn / d\omega]}.
\label{eq2}
\end{equation}

We see from eq.~(\ref{eq2}) that if the index of refraction decreases
rapidly enough with frequency, the group velocity can be negative.
It is well known that the index of refraction decreases rapidly with
frequency near an absorption line, where ``anomalous'' wave propagation 
effects can occur \cite{Brillouin}.  However, the absorption makes it 
difficult to study these effects.   The insight of Garrett and McCumber
\cite{Garrett} and of Chiao \etal\
\cite{Chiao93,Bolda94,Steinberg94,Chiao96,Chiao97} 
is that demonstrations of negative group
velocity are possible in media with inverted 
populations, so that gain rather than absorption occurs at the frequencies of 
interest.
This was dramatically realized in the experiment of 
Wang \etal\ \cite{Wang} by use of a closely spaced pair of gain lines,
as perhaps first suggested by Steinberg and Chiao \cite{Steinberg94}.
 
We use a classical oscillator model for the index of
refraction.  The index $n$ is the square root of the dielectric constant
$\epsilon$, which is in turn related to the atomic polarizability $\alpha$
according to 
\begin{equation}
D = \epsilon E = E + 4 \pi P = E(1 + 4 \pi N \alpha),
\label{eq3}
\end{equation}
(in Gaussian units)
where $D$ is the electric displacement, $E$ is the electric field and
$P$ is the polarization density.  Then, the index of refraction of a dilute
gas is
\begin{equation}
n = \sqrt{\epsilon} \approx 1 + 2 \pi N \alpha.
\label{eq4}
\end{equation}

The polarizability $\alpha$ is obtained from the electric dipole moment 
$p = ex = \alpha E$ induced by electric field $E$.  In the case of a single 
spectral line of frequency $\omega_j$, we say that an electron
is bound to the (fixed) nucleus by a spring of constant
$K = m \omega_j^2$, and that the motion is subject to the damping force
$-m \gamma_j \dot x$, where the dot indicates differentiation with respect to
time.
The equation of motion in the presence of an electromagnetic
wave of frequency $\omega$ is
\begin{equation}
\ddot x + \gamma_j \dot x + \omega_j^2 x = {e E \over m} = {e E_0 \over m}
 e^{i\omega t}.
\label{eq5}
\end{equation}
Hence,
\begin{equation}
x = {e E \over m} {1 \over \omega_j^2 - \omega^2 - i \gamma_j \omega}
= {e E \over m} {\omega_j^2 - \omega^2 + i \gamma_j \omega \over
(\omega_j^2 - \omega^2)^2 + \gamma_j^2 \omega^2},
\label{eq6}
\end{equation}
and the polarizability is
\begin{equation}
\alpha = {e^2\over m} {\omega_j^2 - \omega^2 + i \gamma_j \omega \over
(\omega_j^2 - \omega^2)^2 + \gamma_j^2 \omega^2}.
\label{eq7}
\end{equation}

In the present problem, we have two spectral lines, $\omega_{1,2} = \omega_0 
\pm \Delta/2$, both of oscillator strength $-1$ to indicate
that the populations of both lines are inverted, with 
damping constants $\gamma_1 = \gamma_2 = \gamma$.
In this case, the polarizability is given by
\begin{eqnarray}
\alpha & = & - {e^2\over m} {(\omega_0 - \Delta/2)^2 - \omega^2 
 + i \gamma \omega \over
((\omega_0 - \Delta/2)^2 - \omega^2)^2 + \gamma^2 \omega^2}
- {e^2\over m} {(\omega_0 + \Delta/2)^2 - \omega^2 
+ i \gamma \omega \over
((\omega_0 + \Delta/2)^2 - \omega^2)^2 + \gamma^2 \omega^2}
\nonumber \\
& \approx & - {e^2\over m} {\omega_0^2 - \Delta \omega_0 - 
\omega^2 + i \gamma \omega \over
(\omega_0^2 - \Delta \omega_0 - \omega^2)^2 + \gamma^2 \omega^2}
- {e^2\over m} {\omega_0^2 + 2\Delta \omega_0 - \omega^2 
+ i \gamma \omega \over
(\omega_0^2 + \Delta \omega_0 - \omega^2)^2 + \gamma^2 \omega^2},
\label{eq8}
\end{eqnarray}
where the approximation is obtained by the neglect of 
terms in $\Delta^2$ compared to those in $\Delta \omega_0$.

\begin{figure}[htp]  
\begin{center}
\vspace{0.1in}
\includegraphics[width=4in]{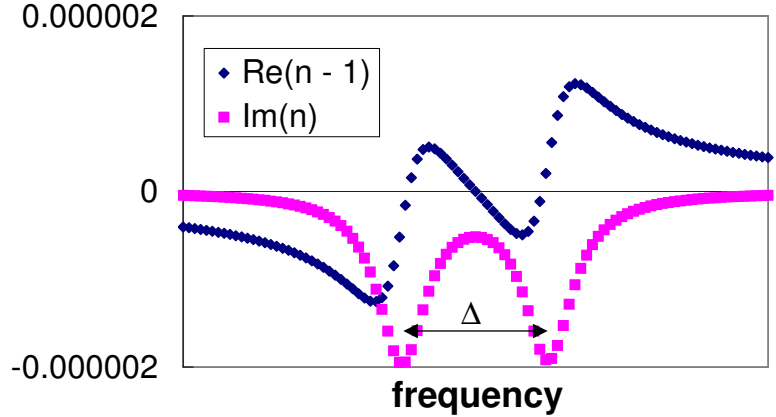}
\parbox{5.5in} 
{\caption[ Short caption for table of contents ]
{\label{fig1} The real and imaginary parts of the index of refraction
in a medium with two spectral lines that have been pumped to inverted 
populations.  The lines are separated by angular frequency $\Delta$ and have
widths $\gamma = 0.4 \Delta$.
}}
\end{center}
\end{figure}

For a probe beam at frequency
$\omega$, the index of refraction (\ref{eq4}) has the form
\begin{equation}
n(\omega) \approx 1 - {\omega_p^2 \over 2} \left[
{\omega_0^2 - \Delta \omega_0 - 
\omega^2 + i \gamma \omega \over
(\omega_0^2 - \Delta \omega_0 - \omega^2)^2 + \gamma^2 \omega^2}
+ {\omega_0^2 + \Delta \omega_0 - \omega^2 
+ i \gamma \omega \over
(\omega_0^2 + \Delta \omega_0 - \omega^2)^2 + \gamma^2 \omega^2}
\right],
\label{eq9}
\end{equation}
where $\omega_p$ is the plasma frequency given by eq.~(\ref{eq0}).
This illustrated in Figure \ref{fig1}.

The index at the central frequency $\omega_0$ is
\begin{equation}
n(\omega_0) \approx 1 - i {\omega_p^2 \gamma \over (\Delta^2 + \gamma^2) 
\omega_0}
\approx 1 - i {\omega_p^2 \over \Delta^2} {\gamma \over \omega_0},
\label{eq10}
\end{equation}
where the second approximation holds when $\gamma \ll \Delta$.
The electric field of a continuous 
probe wave then propagates according to 
\begin{equation}
E(z,t) = e^{i(kz - \omega_0 t)} 
= e^{i \omega_0 (n(\omega_0) z/ c - t)}
\approx e^{z/[\Delta^2 c / \gamma \omega_p^2]} e^{i \omega_0 (z/c - t)}.
\label{eq11}
\end{equation}
From this we see that at frequency $\omega_0$ the phase velocity is $c$,
and the medium has an amplitude gain length $\Delta^2 c / \gamma \omega_p^2$.

\begin{figure}[htp]  
\begin{center}
\vspace{0.1in}
\includegraphics[width=3in]{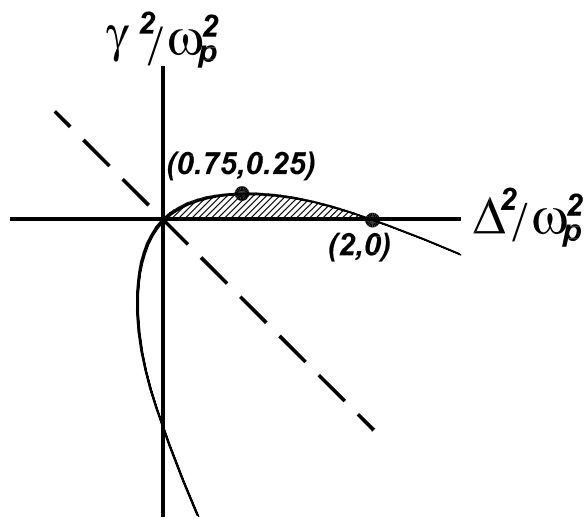}
\parbox{5.5in} 
{\caption[ Short caption for table of contents ]
{\label{allow} The allowed region (\ref{eq13})
in $(\Delta^2,\gamma^2)$ space such that the group velocity is negative.
}}
\end{center}
\end{figure}

To obtain the group velocity (\ref{eq2})
at frequency $\omega_0$, we need the derivative
\begin{equation}
{d(\omega n) \over d\omega} \biggr|_{\omega_0} \approx
1 - { 2 \omega_p^2 (\Delta^2 - \gamma^2) \over (\Delta^2 + \gamma^2)^2},
\label{eq12}
\end{equation}
where we have neglected terms in $\Delta$ and $\gamma$ compared to
$\omega_0$.
From eq.~(\ref{eq2}), we see that the group velocity can be
negative if
\begin{equation}
{\Delta^2 \over \omega_p^2} - {\gamma^2 \over \omega_p^2} \ge {1 \over 2}
\left( {\Delta^2 \over \omega_p^2} + {\gamma^2 \over \omega_p^2} \right)^2.
\label{eq13}
\end{equation} 
The boundary of the allowed region (\ref{eq13})
in $(\Delta^2,\gamma^2)$ space is a
parabola whose axis is along the line $\gamma^2 = - \Delta^2$, as shown in
Fig.~\ref{allow}.  For the physical region $\gamma^2 \ge 0$, the boundary
is given by
\begin{equation}
{\gamma^2 \over \omega_p^2} = \sqrt{1 + 4 {\Delta^2 \over \omega_p^2}} - 1
- {\Delta^2 \over \omega_p^2}.
\label{eq13b}
\end{equation}
Thus, to have a negative group velocity, we must have
\begin{equation}
\Delta \le \sqrt{2} \omega_p,
\label{eq13c}
\end{equation}
which limit is achieved when $\gamma = 0$; the maximum value of $\gamma$ is
$0.5 \omega_p$ when $\Delta = 0.866 \omega_p$.

Near the boundary of the negative group velocity region, $\abs{v_g}$ exceeds 
$c$, which alerts us to concerns of superluminal behavior.  However, as will
be seen in the following sections, the effect of a negative group velocity is
more dramatic when $\abs{v_g}$ is small rather than large.

The region of recent experimental interest is 
$\gamma \ll \Delta \ll \omega_p$, for which eqs.~(\ref{eq2}) and (\ref{eq12})
predict that
\begin{equation}
v_g \approx - {c \over 2} {\Delta^2 \over \omega_p^2}.
\label{eq13a}
\end{equation}
A value of  $v_g \approx - c / 310$ as in the
experiment of Wang corresponds to $\Delta / \omega_p \approx 1/12$.  In this
case, the gain length $\Delta^2 c / \gamma \omega_p^2$ was approximately
40 cm.

For later use we record the second derivative,
\begin{equation}
{d^2(\omega n) \over d\omega^2} \biggr|_{\omega_0} \approx 8 i
{\omega_p^2 \gamma (3 \Delta^2 - \gamma^2) \over 
(\Delta^2 + \gamma^2)^3} 
\approx 24 i {\omega_p^2 \over \Delta^2} {\gamma \over \Delta^2}
\label{eq14}
\end{equation}
where the second approximation holds if $\gamma \ll \Delta$.

\subsection{Propagation of a Monochromatic Plane Wave}

To illustrate the optical properties of a medium with negative group
velocity, we consider the propagation of an electromagnetic wave through
it.  The medium extends from $z = 0$ to $a$, and is surrounded by vacuum.
Because the index of refraction (\ref{eq9}) is near unity in the
frequency range of interest, we ignore reflections at the boundaries of
the medium.  

A monochromatic plane wave of frequency $\omega$
and incident from $z < 0$ propagates with phase velocity $c$ in
vacuum.  Its electric field can be written
\begin{equation}
E_\omega(z,t) = E_0 e^{i \omega z/c} e^{-i \omega t} \qquad (z < 0).
\label{eq15}
\end{equation}
Inside the medium this wave propagates with phase velocity $c / n(\omega)$
according to
\begin{equation}
E_\omega(z,t) = E_0 e^{i \omega n z/c} e^{-i \omega t} \qquad 
(0 < z < a),
\label{eq16}
\end{equation}
where the amplitude is unchanged since we neglect the small reflection at
the boundary $z = 0$.
When the wave emerges into vacuum at $z = a$, the phase velocity is again
$c$, but it has accumulated a phase lag
of $(\omega / c)(n - 1)a$, and so appears as
\begin{equation}
E_\omega(z,t) = E_0 e^{i\omega a (n - 1) / c} e^{i \omega z/c}  e^{-i \omega t}
= E_0 e^{i \omega a n / c} e^{-i \omega (t - (z - a)/c)}
 \qquad (a < z).
\label{eq17}
\end{equation}
It is noteworthy that a monochromatic wave for $z > a$ has the same form as 
that inside the medium if we make the frequency-independent substitutions 
\begin{equation}
z \to a, \qquad \mbox{and} \qquad t \to t - {z - a \over c}.
\label{eq17a}
\end{equation}

Since an arbitrary waveform can be expressed in terms of monochromatic
plane waves via Fourier analysis, we can use these substitutions to convert
any wave in the region $0 < z < a$ to its continuation in the region $a < z$.

A general relation can be deduced in the case where the second and higher
derivatives of $\omega n(\omega)$ are very small.  We can then write 
\begin{equation}
\omega n(\omega) \approx \omega_0 n(\omega_0)
+ {c \over v_g} (\omega - \omega_0),
\label{eq17b}
\end{equation}
where $v_g$ is the group velocity for a pulse with central frequency
$\omega_0$.  Using this in eq.~(\ref{eq16}), we have
\begin{equation}
E_\omega(z,t) \approx E_0 e^{i \omega_0 z(n(\omega_0) /c - 1/v_g)} 
e^{i \omega z / v_g} e^{-i \omega t} \qquad 
(0 < z < a).
\label{eq17c}
\end{equation}
In this approximation, the Fourier component $E_\omega(z)$ 
at frequency $\omega$ of a wave
inside the gain medium is related to that of the incident wave by replacing 
the frequency dependence $e^{i \omega z / c}$ by $e^{i \omega z / v_g}$,
\ie, by replacing $z/c$ by $z/v_g$, and
multiplying by the frequency-independent phase factor 
$e^{i \omega_0 z(n(\omega_0) /c - 1/v_g)}$.  Then, using transformation
(\ref{eq17a}), the wave that emerges into vacuum beyond the medium is
\begin{equation}
E_\omega(z,t) \approx E_0 e^{i \omega_0 a(n(\omega_0) /c - 1/v_g)} 
e^{i \omega (z / c - a(1 / c - 1 / v_g))} e^{-i \omega t} \qquad 
(a < z).
\label{eq17e}
\end{equation}
The wave beyond the medium is related to the incident wave by multiplying
by a frequency-independent phase, and by replacing $z/c$ by 
$z / c - a(1 / c - 1 / v_g)$ in the frequency-dependent part of the phase.

The effect of the medium on the wave as described by 
eqs.~(\ref{eq17c})-(\ref{eq17e}) has been called ``rephasing" \cite{Wang}.

\subsection{Fourier Analysis and ``Rephasing''}

The transformations between the monochromatic incident wave (\ref{eq15})
and its continuation in and
beyond the medium, (\ref{eq17c}) and (\ref{eq17e}),
imply that an incident wave
\begin{equation}
E(z,t) = f(z/c - t) = \int_{-\infty}^\infty d\omega\, E_\omega(z) 
e^{-i \omega t} \qquad (z < 0),
\label{s3}
\end{equation}
whose Fourier components are given by
\begin{equation}
E_\omega(z) = {1 \over 2 \pi} \int_{-\infty}^\infty dE(z,t) e^{i \omega t} 
dt,
\label{s4}
\end{equation}
propagates as
\begin{eqnarray}
E(z,t) \approx \left\{ \begin{array}{ll}
 f(z/c - t) & \qquad (z < 0), \\ 
e^{i \omega_0 z (n(\omega_0)/c - 1/v_g)}
f(z/v_g - t) & \qquad (0 < z < a), \\ 
e^{i \omega_0 a (n(\omega_0)/c - 1/v_g)}
f(z/c - t - a(1/c - 1/v_g)) & \qquad (a < z).
\end{array} \right.
\label{eq17d}
\end{eqnarray}

An interpretation of eq.~(\ref{eq17d}) in terms of ``rephasing'' is as follows.
Fourier analysis tells us that the maximum amplitude of a
pulse made of waves of many frequencies, each of the form $E_\omega(z,t)
= E_0(\omega) e ^{i \phi(\omega)} 
= E_0(\omega) e ^{i (k(\omega) z - \omega t + \phi_0(\omega))}$
with $E_0 \ge 0$, is determined by adding the amplitudes $E_0(\omega)$.
This maximum is achieved only if there exists points $(z,t)$ such that all
phases $\phi(\omega)$ have the same value.

\begin{figure}[htp]  
\begin{center}
\vspace{0.1in}
\includegraphics[width=6in]{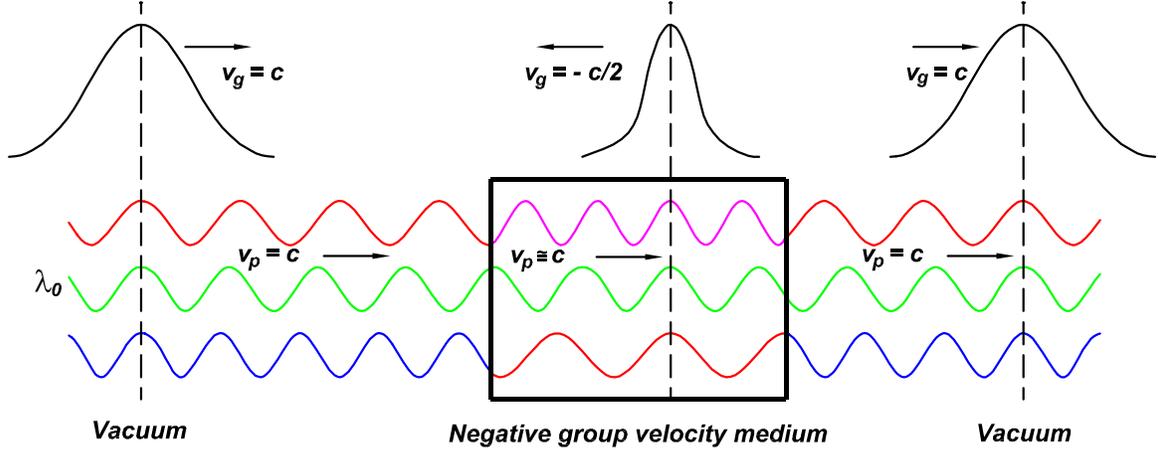}
\parbox{5.5in} 
{\caption[ Short caption for table of contents ]
{\label{rephase} A snapshot of three Fourier components of a pulse in
the vicinity of a negative group velocity medium.  The component at the central
 wavelength $\lambda_0$ is unaltered by the medium, but the wavelength
of a longer wavelength component is shortened, and that of a shorter
wavelength component is lengthened.  Then, even when the incident pulse has not
yet reached the medium, there can be a point inside the
medium at which all components have the same phase, and a peak appears.  
Simultaneously, there can be a point in the vacuum region beyond the medium at 
which the Fourier components are again all in phase, and a third peak appears.  
The peaks in the vacuum regions move with group 
velocity $v_g = c$, but the peak inside the medium moves
with a negative group velocity, shown as $v_g = - c/2$ in the
figure.  The phase velocity $v_p$ is $c$ in vacuum, and close to $c$ in the
medium.
}}
\end{center}
\end{figure}

For example, we consider a pulse in the region $z < 0$ whose maximum occurs
when the phases of all component frequencies vanish, as shown at the left
of Fig.~\ref{rephase}.  Referring to eq.~(\ref{eq15}),
we see that the peak occurs when $z = c t$.  As usual, we say that the
group velocity of this wave is $c$ in vacuum.
  
Inside the medium, eq.~(\ref{eq17c}) describes the phases of the components, 
which all have a common frequency-independent
phase $\omega_0 z(n(\omega_0) /c - 1/v_g)$ at a given $z$, as
well as a frequency-dependent part $\omega (z / v_g - t)$.  The peak of the
pulse occurs when all the frequency-dependent phases vanish; the overall
frequency-independent phase does not affect the pulse size.  Thus, the
peak of the pulse propagates within the medium according to $z = v_g t$.
The velocity of the peak is $v_g$, the group velocity of the medium, which
can be negative.  

The ``rephasing'' (\ref{eq17c}) within the medium changes the wavelengths
of the component waves.  Typically the wavelength increases, and by
greater amounts at longer wavelengths.  A longer time is required 
before the phases of the waves all becomes the same at some point $z$ inside
the medium, so in a normal medium the velocity of the peak appears to be
slowed down.  But in a negative group velocity medium,
wavelengths short compared to $\lambda_0$ are lengthened, long waves
are shortened, and the velocity of the peak appears to be reversed.

By a similar argument, eq.~(\ref{eq17e}) tells us that in the vacuum region
beyond the medium  
the peak of the pulse propagates according to $z = ct + a(1 / c - 1 / v_g)$.
The group velocity is again $c$, but the ``rephasing'' within the medium
results in a shift of the position of the peak by amount 
$a(1 / c - 1 / v_g)$.  In a normal medium where $0 < v_g \leq c$ the shift
is negative; the pulse appears to have been delayed during its
 passage through the medium.
But after a negative group velocity medium, the pulse appears to have advanced!

This advance is possible because in the Fourier view, each component wave
extends over all space, even if the pulse appears to be restricted.  The
unusual ``rephasing'' in a negative group velocity medium shifts the phases
of the frequency components of the wave train in the region ahead of the
nominal peak such that the phases all coincide, and a peak is observed,
at times earlier than expected at points beyond the medium.

As shown in Fig.~\ref{rephase} and further illustrated in the examples below, 
the ``rephasing'' can result in
the simultaneous appearance of peaks in all three regions.

\subsection{Propagation of a Sharp Wave Front}

To assess the effect of a medium with negative group velocity on the 
propagation of a signal, we first consider a waveform with a sharp front,
as recommended by Sommerfeld and Brillouin \cite{Brillouin}.

As an extreme but convenient example, we take the
incident pulse to be a Dirac delta function, $E(z,t) = E_0 \delta(z/c - t)$.
Inserting this in eq.~(\ref{eq17d}), which is based on the linear
approximation (\ref{eq17b}), we find
\begin{eqnarray}
E(z,t) \approx \left\{ \begin{array}{ll}
 E_0 \delta(z/c - t) & \qquad (z < 0), \\ 
E_0 e^{i \omega_0 z (n(\omega_0)/c - 1/v_g)}
\delta (z/v_g - t) & \qquad (0 < z < a), \\ 
E_0 e^{i \omega_0 a (n(\omega_0) / c - 1 / v_g)} 
\delta(z / c - t - a(1/c - 1/v_g)) & \qquad (a < z),
\end{array} \right.
\label{eq35}
\end{eqnarray}

According to eq.~(\ref{eq35}), the delta-function pulse emerges from the
medium at $z = a$ at time $t = a/v_g$.  If the group velocity is 
negative, the pulse emerges from the medium before it enters at $t = 0$!

A sample history of (Gaussian) 
pulse propagation is illustrated in Fig.~\ref{fig2}.
Inside the negative group velocity medium, an (anti)pulse propagates backwards
in space from $z = a$ at time $t = a / v_g < 0$ to $z = 0$ at time $t = 0$, 
at which point it appears to annihilate the incident pulse.

This behavior is
analogous to barrier penetration by a relativistic electron \cite{Feynman}
in which an electron can emerge from the far side of the barrier earlier
than it hits the near side, if the electron emission at the far side is
accompanied by positron emission, and the positron propagates within the 
barrier so as to annihilate the incident electron at the near side.
In the Wheeler-Feynman
view, this process involves only a single electron which 
propagates backwards in time when inside the barrier.  In this spirit,
we might say that pulses propagate backwards in time (but forward in space)
inside a negative group velocity medium.

The Fourier components of the delta function are independent of frequency,
so the advanced appearance of the sharp wavefront as described by 
eq.~(\ref{eq35}) can occur only for a gain medium such that the index of refraction varies linearly at all frequencies.  If such a medium existed
with negative slope $dn/d\omega$, then  eq.~(\ref{eq35}) would
constitute superluminal signal propagation.

However, from Fig.~\ref{fig1} we see that a linear approximation to the
index of refraction is reasonable in the negative group velocity
 medium only for $\abs{\omega - \omega_0}
\lsim \Delta/2$.  The sharpest wavefront that can be supported within this
bandwidth has characteristic risetime $\tau \approx 1 /\Delta$.

For the experiment of Wang \etal\ where $\Delta/2 \pi \approx 10^6$ Hz, an
analysis based on eq.~(\ref{eq17b}) would be valid only for pulses
with $\tau \gsim 0.1\ \mu$s.  Wang \etal\ used a pulse with $\tau
\approx 1\ \mu$s, close to the minimum value for which eq.~(\ref{eq17b}) is a
reasonable approximation. 

Since a negative group velocity can only be experienced over a limited
bandwidth, very sharp wavefronts must be excluded from discussion of
signal propagation.  However, it is well known \cite{Brillouin}
that great care must be taken when discussing the signal velocity if the
waveform is not sharp.


\subsection{Propagation of a Gaussian Pulse}

We now consider a Gaussian pulse of temporal length $\tau$ centered on
frequency $\omega_0$ (the carrier frequency), for which the incident
waveform is
\begin{equation}
E(z,t) =  E_0 e^{-(z/c - t)^2 / 2 \tau^2} e^{i \omega_0 z/c} 
e^{-i \omega_0 t} \qquad (z < 0),
\label{eq18}
\end{equation}
Inserting this in eq.~(\ref{eq17d}) we find
\begin{eqnarray}
E(z,t) = \left\{ \begin{array}{ll}
E_0 e^{-(z/c - t)^2 / 2 \tau^2} e^{i \omega_0 (z/c - t)}
& \qquad (z < 0), \\ 
E_0 e^{-(z / v_g - t)^2 / 2 \tau^2} e^{i \omega_0 (n(\omega_0) z/c - t)} 
& \qquad (0 < z < a), \\ 
E_0 e^{i\omega_0 a (n(\omega_0) - 1) / c} 
e^{-(z / c - a(1/c - 1/v_g) - t)^2 / 2 \tau^2} e^{i \omega_0 (z / c - t)} 
& \qquad (a < z).
\end{array} \right.
\label{eq18c}
\end{eqnarray}

The factor $e^{i\omega_0 a (n(\omega_0) - 1) / c}$ in eq.~(\ref{eq18c}) for
$a < z$  becomes
$e^{\omega_p^2 \gamma a / \Delta^2 c}$ using eq.~(\ref{eq10}), and represents a
small gain due to traversing the negative group velocity medium.  
In the experiment of Wang \etal\ this factor was only 1.16.

We have already noted in the previous section that the linear approximation to
$\omega n(\omega)$ is only good over a frequency interval about $\omega_0$
of order $\Delta$, and so eq.~(\ref{eq18c}) for the pulse after the
gain medium applies only for pulsewidths
\begin{equation}
\tau \gsim {1 \over \Delta}.
\label{eq18d}
\end{equation}

Further constraints on the validity of eq.~(\ref{eq18c})
can obtained using the
expansion of $\omega n(\omega)$ to second order.  For this, we repeat the
derivation of eq.~(\ref{eq18c}) in slightly more detail.
The incident Gaussian pulse (\ref{eq18}) has the Fourier decomposition 
(\ref{s4})
\begin{equation}
E_\omega(z) 
= {\tau \over \sqrt{2 \pi}} E_0 e^{- \tau^2 (\omega - \omega_0)^2 / 2}
e^{i \omega z / c} \qquad (z < 0).
\label{eq20}
\end{equation}
We again extrapolate the Fourier component at frequency $\omega$ into
the region $z > 0$ using eq.~(\ref{eq16}),
which yields
\begin{equation}
E_\omega(z) 
= {\tau \over \sqrt{2 \pi}} E_0 e^{- \tau^2 (\omega - \omega_0)^2 / 2}
e^{i \omega n z / c} \qquad (0 < z < a).
\label{eq21}
\end{equation}

We now approximate the factor
$\omega n(\omega)$ by its Taylor expansion through second order:
\begin{equation}
\omega n(\omega) \approx \omega_0 n(\omega_0)
+ {c \over v_g} (\omega - \omega_0)
+ {1 \over 2} {d^2(\omega n) \over d\omega^2 } \Biggr|_{\omega_0} 
(\omega - \omega_0)^2.
\label{eq23}
\end{equation}
With this, we find from eqs.~(\ref{s3}) and (\ref{eq21}) that
\begin{equation}
E(z,t) 
= {E_0 \over A} 
e^{-(z / v_g - t)^2 / 2 A^2 \tau^2} e^{i \omega_0 n(\omega_0) z/c} 
e^{-i \omega_0 t} \qquad (0 < z < a),
\label{eq24}
\end{equation}
where
\begin{equation}
A^2(z) = 1 - i{z \over c \tau^2} {d^2(\omega n) \over d\omega^2} 
\Biggr|_{\omega_0}.
\label{eq25}
\end{equation}
The waveform for $z > a$ is obtained from that for $0 < z < a$ by 
the substitutions (\ref{eq17a}) with the result
\begin{equation}
E(z,t) =  {E_0 \over A} e^{i\omega_0 a(n(\omega_0) - 1) / c}
e^{-(z / c - a(1/c - 1/v_g) - t)^2
/ 2 A^2 \tau^2} e^{i \omega_0 z / c} 
e^{-i \omega_0 t} \qquad (a < z),
\label{eq27}
\end{equation}
where $A$ is evaluated at $z = a$ here.  As expected, the forms (\ref{eq24}) 
and (\ref{eq27}) revert to those of eq.~(\ref{eq18c}) when 
$d^2(\omega n(\omega_0)) / d\omega^2 = 0$.

So long as the factor $A(a)$ is not greatly different from unity, the pulse
emerges from the medium essentially undistorted, which requires
\begin{equation}
{a \over c \tau} \ll {1 \over 24} {\Delta^2 \over \omega_p^2}
{\Delta \over \gamma} \Delta \tau,
\label{eq26}
\end{equation}
using eqs.~(\ref{eq14}) and (\ref{eq25}).
In the experiment of Wang \etal, this condition is that $a / c\tau \ll 1/120$,
which was well satisfied with $a = 6$ cm and $c \tau = 300$ m.

As in the case of the delta function, the centroid of a Gaussian pulse emerges
from a negative group velocity medium at time 
\begin{equation}
t = {a \over v_g} < 0,
\label{eq28}
\end{equation}
which is earlier than the time $t = 0$ when the centroid enters the medium.
In the experiment of Wang \etal, the time advance of the pulse 
was $a/\abs{v_g} \approx 300 a/ c \approx 6 \times 10^{-8}$ s $\approx
0.06 \tau$.

If one attempts to observe the negative group velocity pulse inside the medium,
the incident wave would be perturbed and the backwards-moving pulse would not
be detected.  Rather, one must deduce the effect of the negative group velocity
medium by observation of the pulse that emerges into the region $z > a$ beyond
that medium, where the significance of the time advance (\ref{eq28}) is the
main issue.

The time advance caused by a negative group velocity 
medium is larger when $\abs{v_g}$ is smaller.  It is possible that
$\abs{v_g} > c$, but this gives a smaller time advance than when the negative
group velocity is such that $\abs{v_g} < c$. 
Hence, there is no special
concern as to the meaning of negative group velocity when $\abs{v_g} > c$.

The maximum possible time advance $t_{\rm max}$ by this technique can be 
estimated from eqs.~(\ref{eq13a}), (\ref{eq26}) and (\ref{eq28}) as 
\begin{equation}
{t_{\rm max} \over \tau} \approx {1 \over 12} {\Delta \over \gamma} 
\Delta \tau \approx 1.
\label{eq28a}
\end{equation}
The pulse can advance by at most a few risetimes due to passage
through the negative group velocity medium.

While this aspect of the pulse propagation appears to be superluminal,
it does not imply superluminal signal propagation.

In accounting for signal propagation time, the time needed to
generate the signal must be included as well.  A pulse with a finite 
frequency bandwidth $\Delta$ takes at least time $\tau \approx 1 /
\Delta$ to be generated, and so is delayed by a time of order
its risetime $\tau$ compared to the case of an idealized sharp wavefront.  
Thus, the advance of a pulse front in a negative group velocity medium by
$\lsim \tau$  can
at most compensate for the original delay in generating that pulse. The
signal velocity, as defined by the path length between the source and
detector divided by the overall time from onset of signal generation to
signal detection,
remains bounded by $c$.

As has been emphasized by Garrett and McCumber \cite{Garrett} and by
Chiao \cite{Chiao96,Chiao97}, the time advance of a pulse emerging from
a gain medium is possible because the forward tail of a smooth pulse gives
advance warning of the later arrival of the peak.  The leading edge of the
pulse can be amplified by the gain medium, which gives the appearance of
superluminal pulse velocities.  However, the medium is merely using 
information stored in the early part of the pulse during its (lengthy)
time of generation to bring the apparent velocity of the pulse closer to $c$.

\begin{figure}[htp]  
\begin{center}
\includegraphics*[width=2.75in, bb = 56 355 531 488]{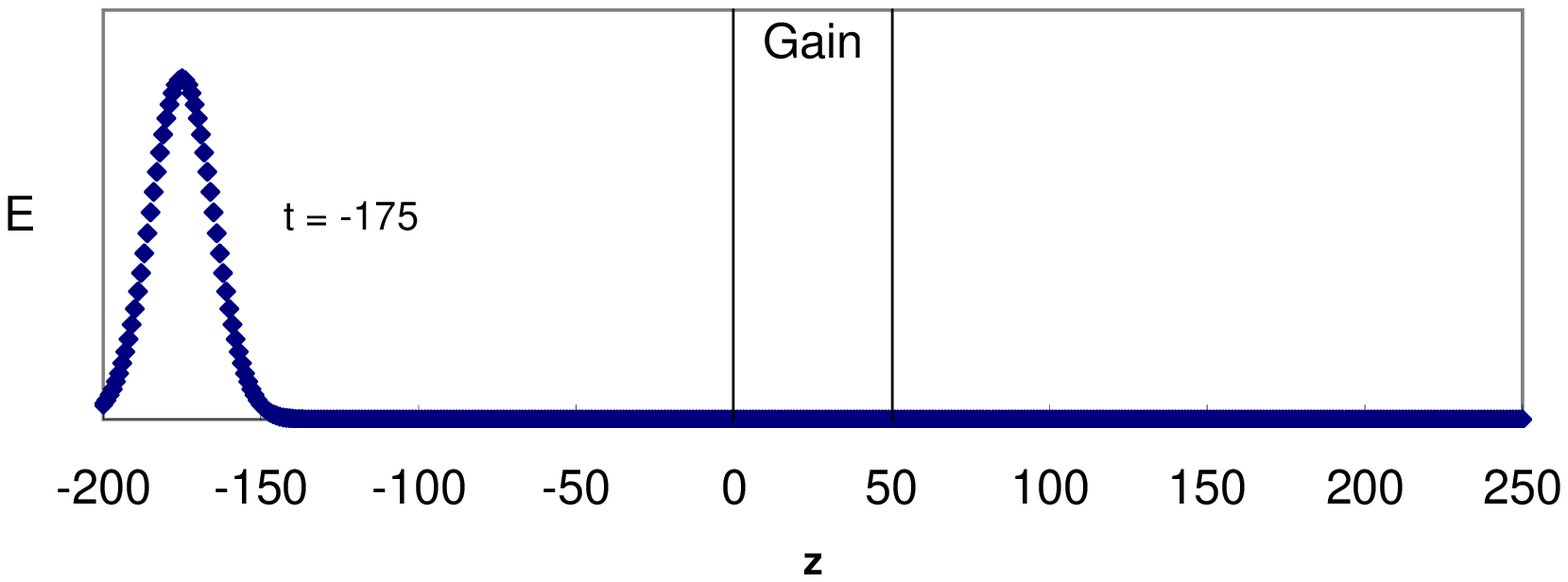} \\
\includegraphics*[width=2.75in, bb = 56 355 531 486]{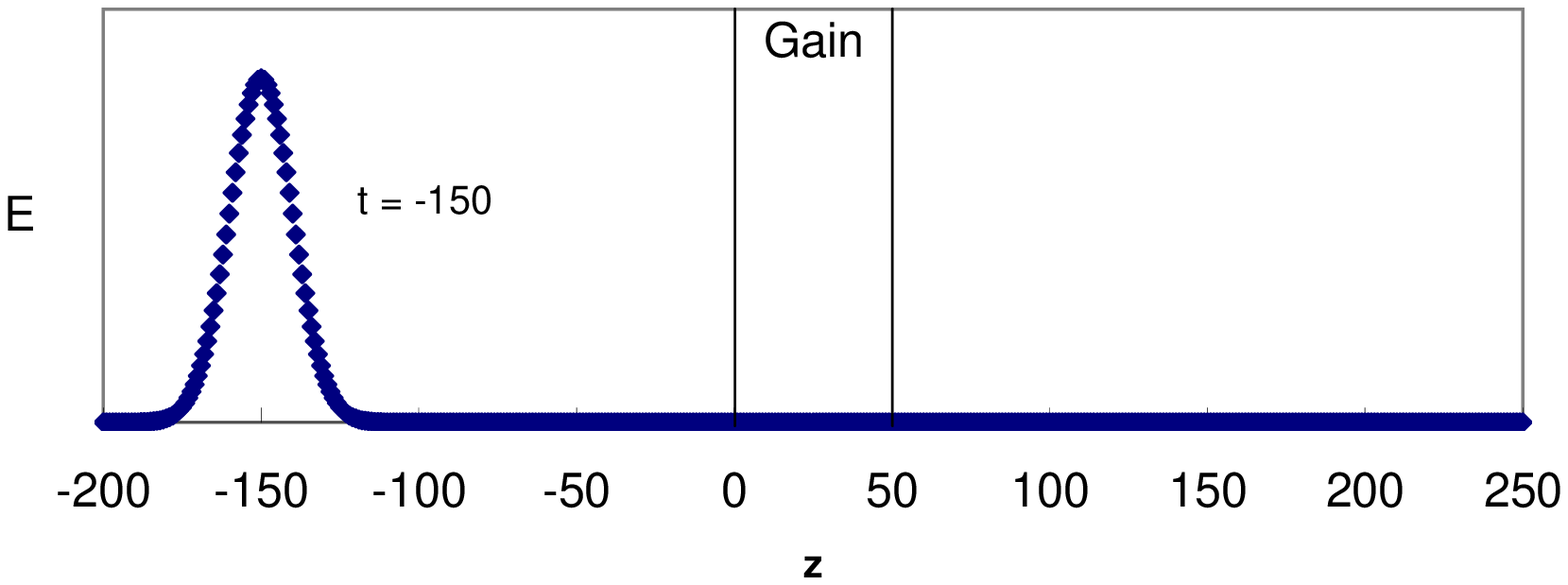} \\
\includegraphics*[width=2.75in, bb = 56 355 531 487]{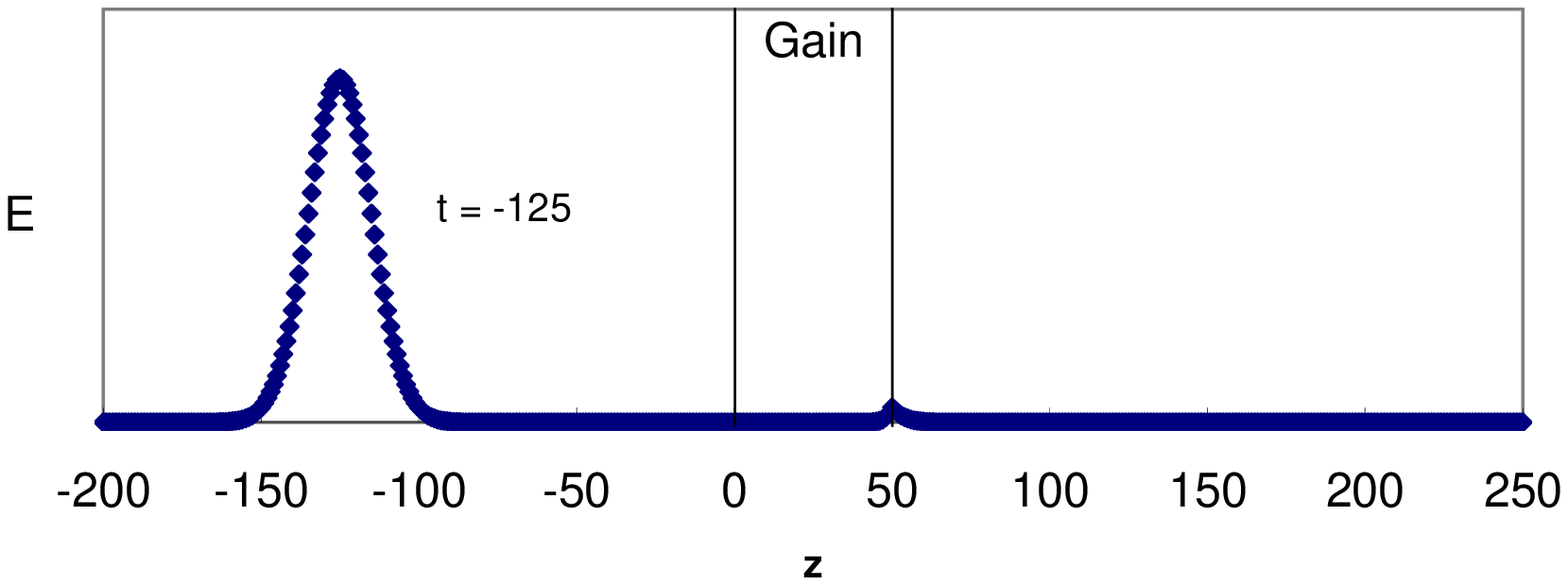} \\
\includegraphics*[width=2.75in, bb = 56 355 531 487]{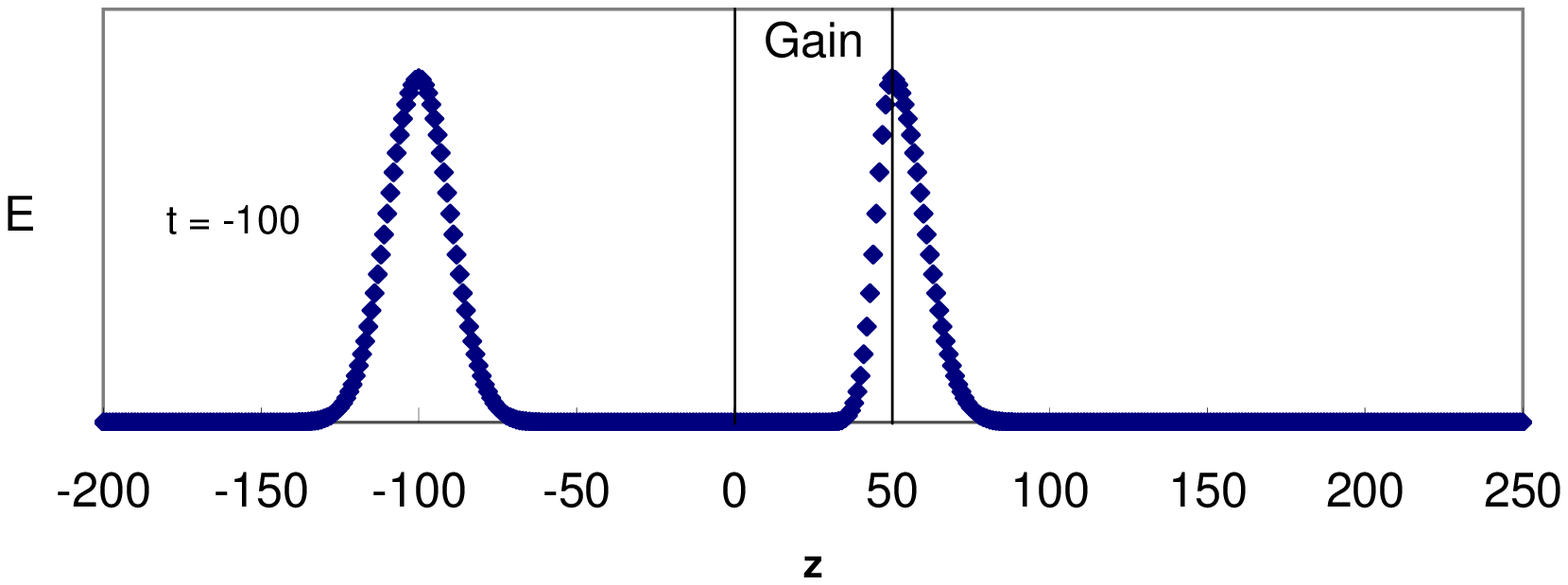} \\
\includegraphics*[width=2.75in, bb = 56 355 531 488]{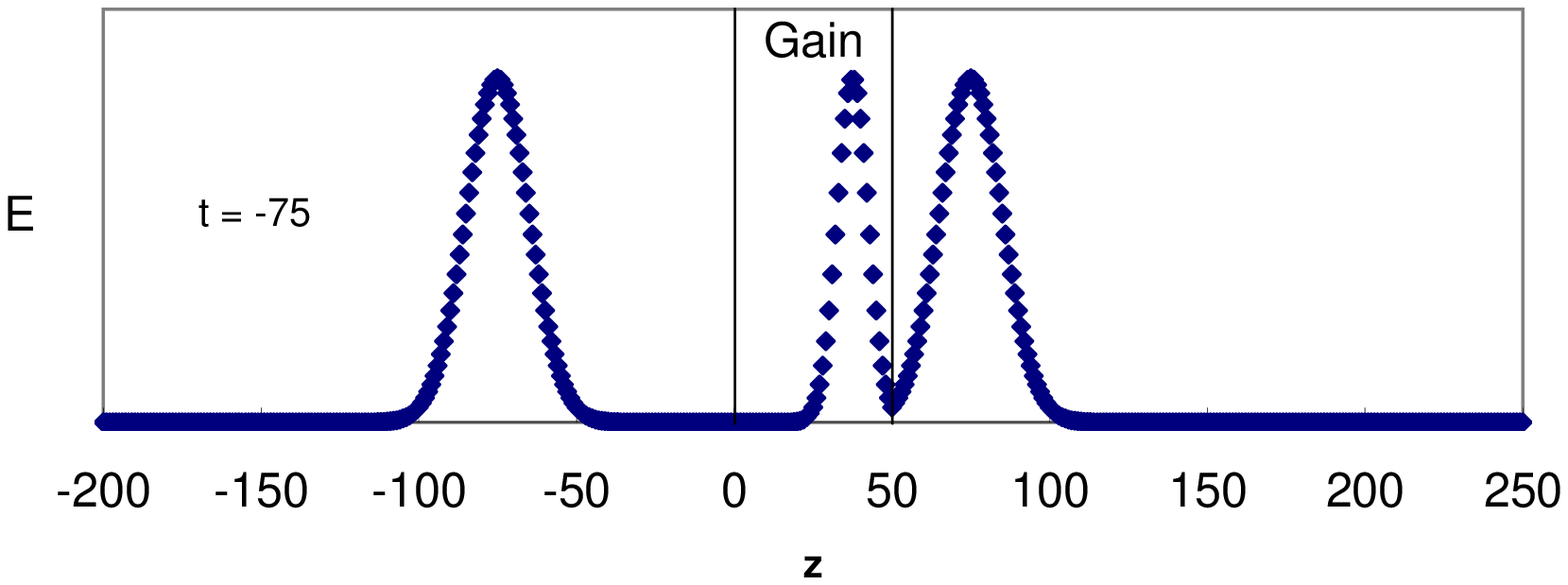} \\
\includegraphics*[width=2.75in, bb = 56 355 531 488]{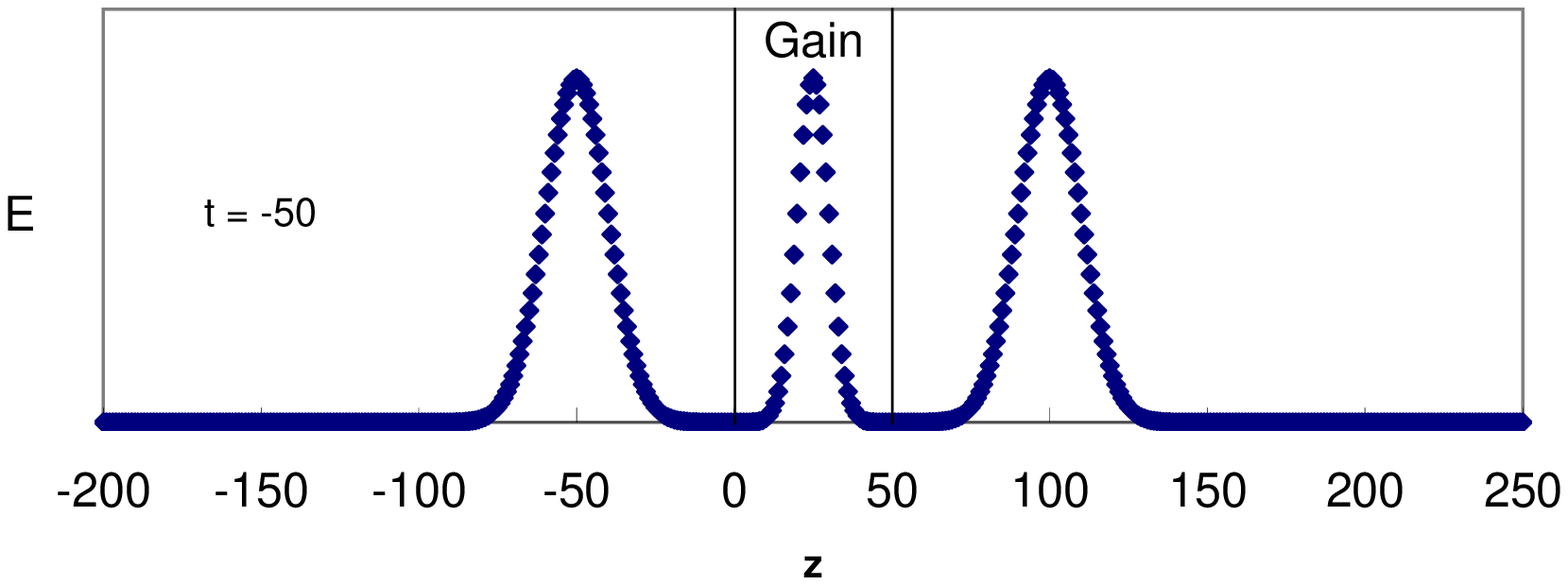} \\
\includegraphics*[width=2.75in, bb = 56 355 531 488]{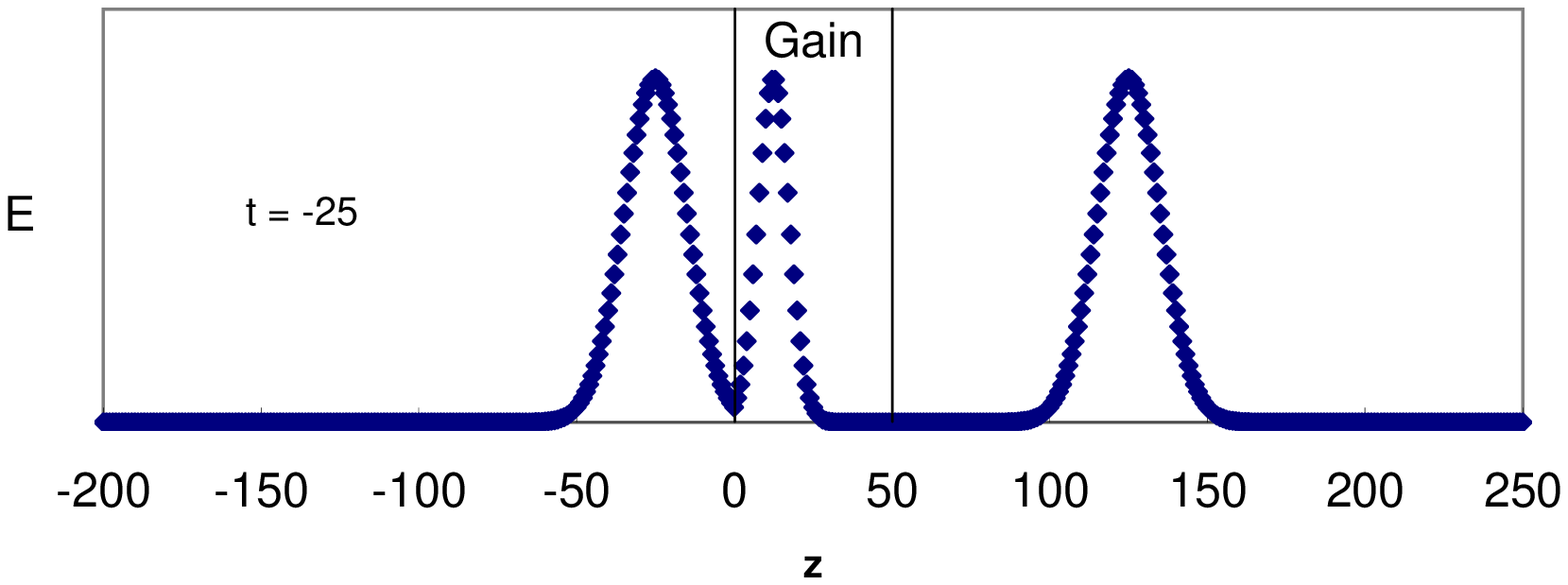} \\
\includegraphics*[width=2.75in, bb = 56 355 531 489]{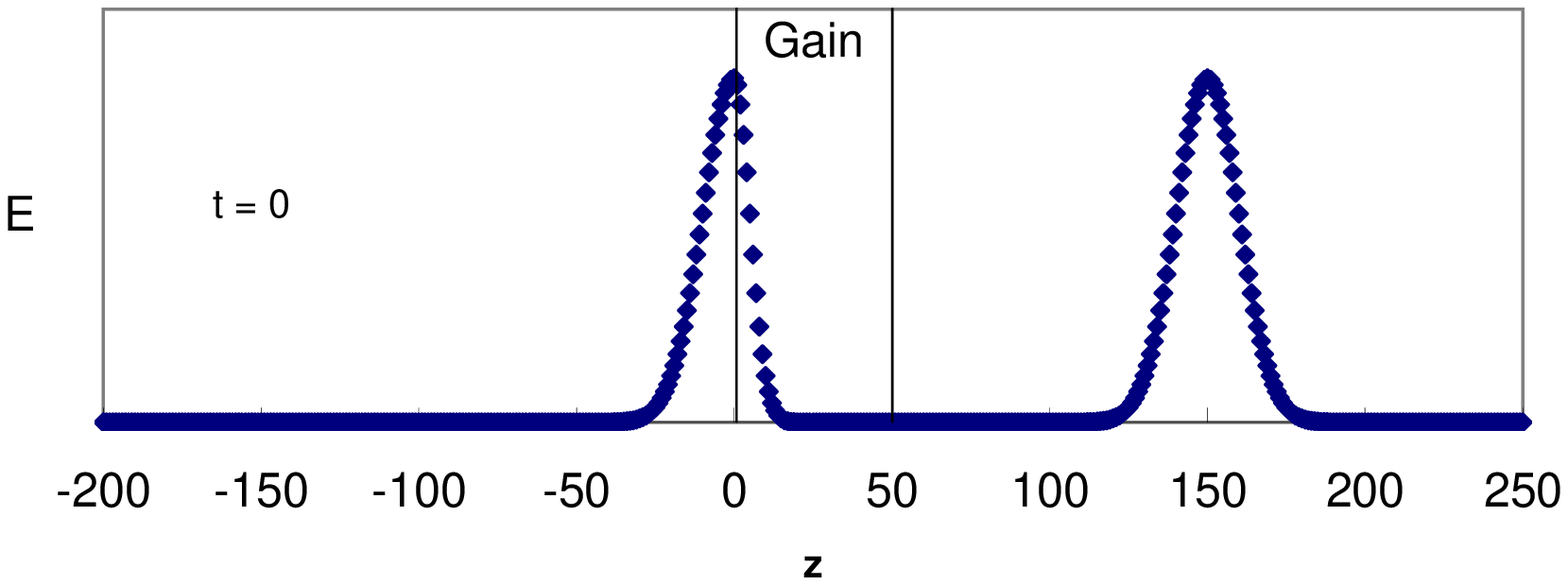} \\
\includegraphics*[width=2.75in, bb = 56 355 531 488]{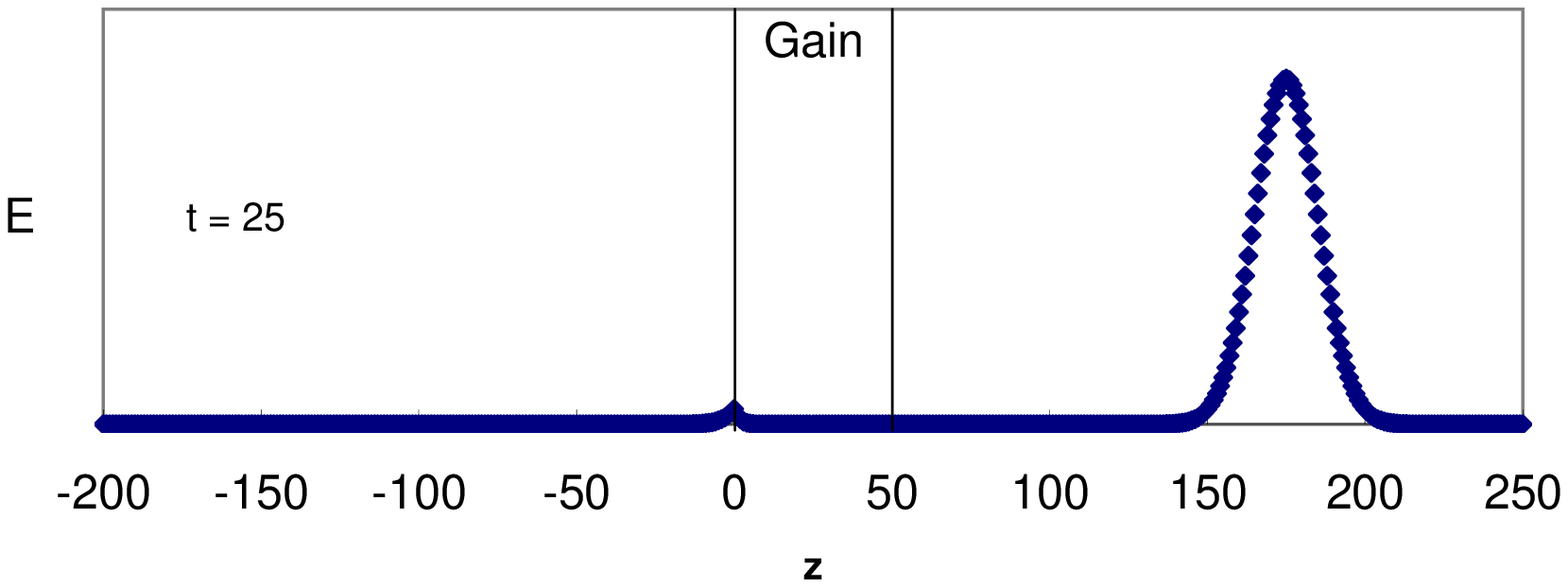} \\
\includegraphics*[width=2.75in, bb = 56 305 531 488]{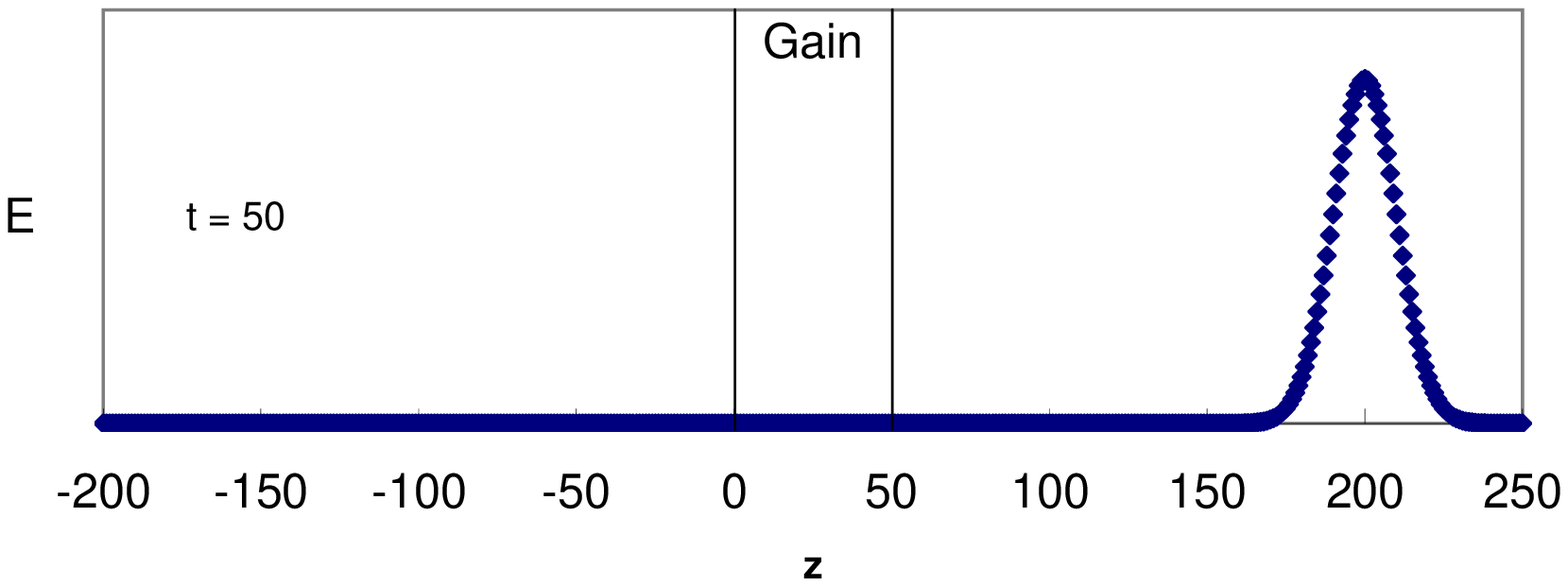}
\parbox{5.5in} 
{\caption[ Short caption for table of contents ]
{\label{fig2} Ten ``snapshots'' of a Gaussian pulse as it traverses a 
negative group velocity region $(0 < z < 50)$, according to 
eq.~(\ref{eq18c}).  The group velocity in the gain medium is $v_g = - c / 2$, 
and $c$ has been set to 1.
}}
\end{center}
\end{figure}

\begin{figure}[htp]  
\begin{center}
\includegraphics*[width=2.75in, bb = 56 355 531 488]{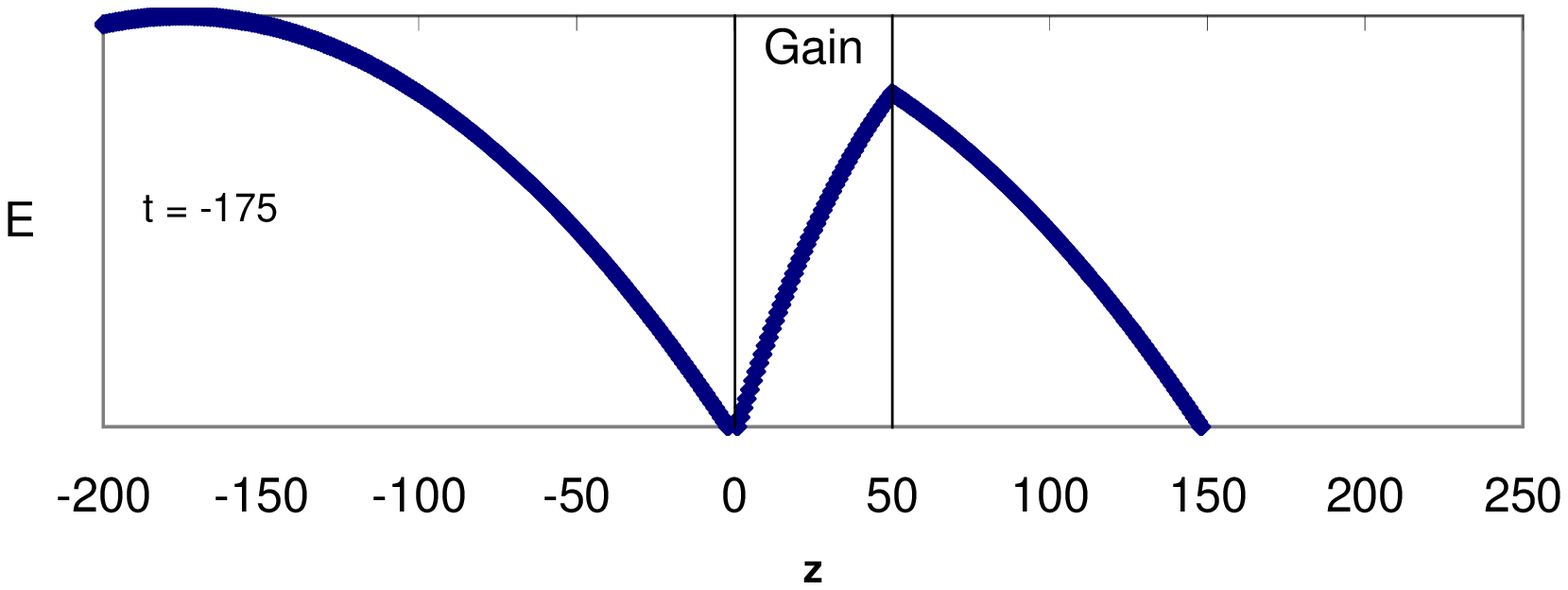} \\
\includegraphics*[width=2.75in, bb = 56 355 531 486]{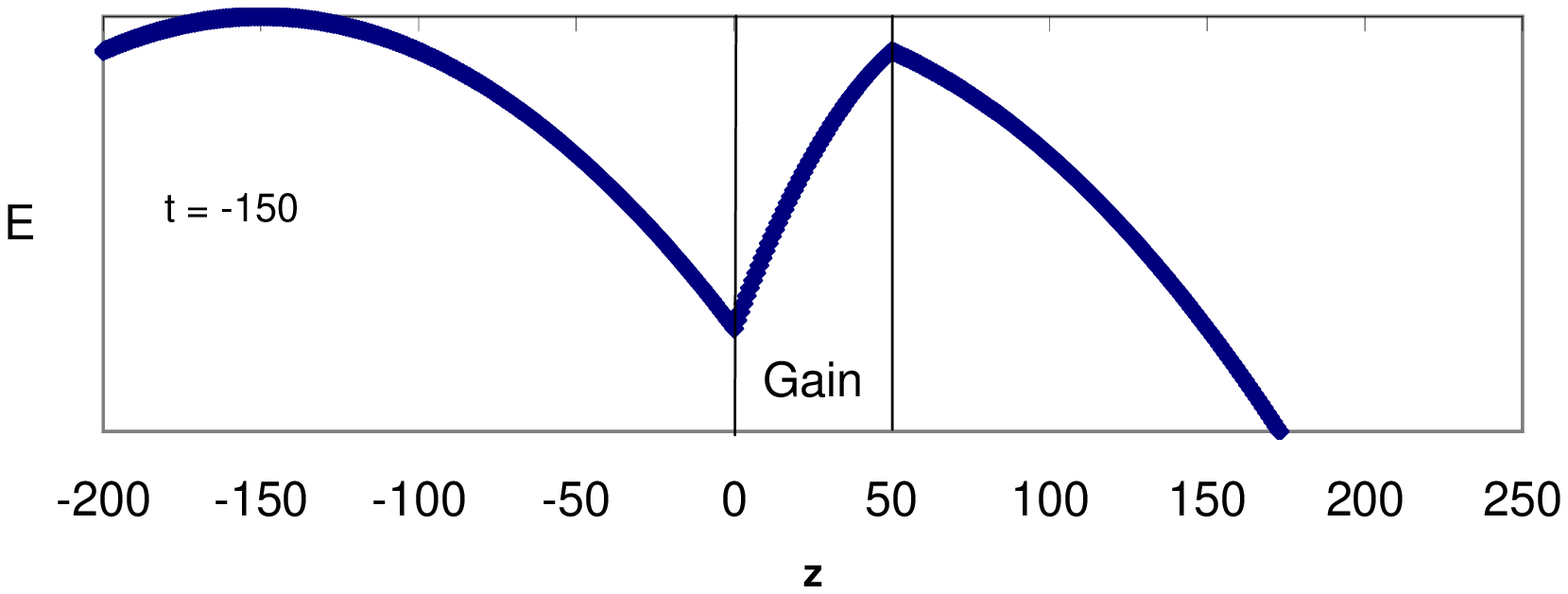} \\
\includegraphics*[width=2.75in, bb = 56 355 531 486]{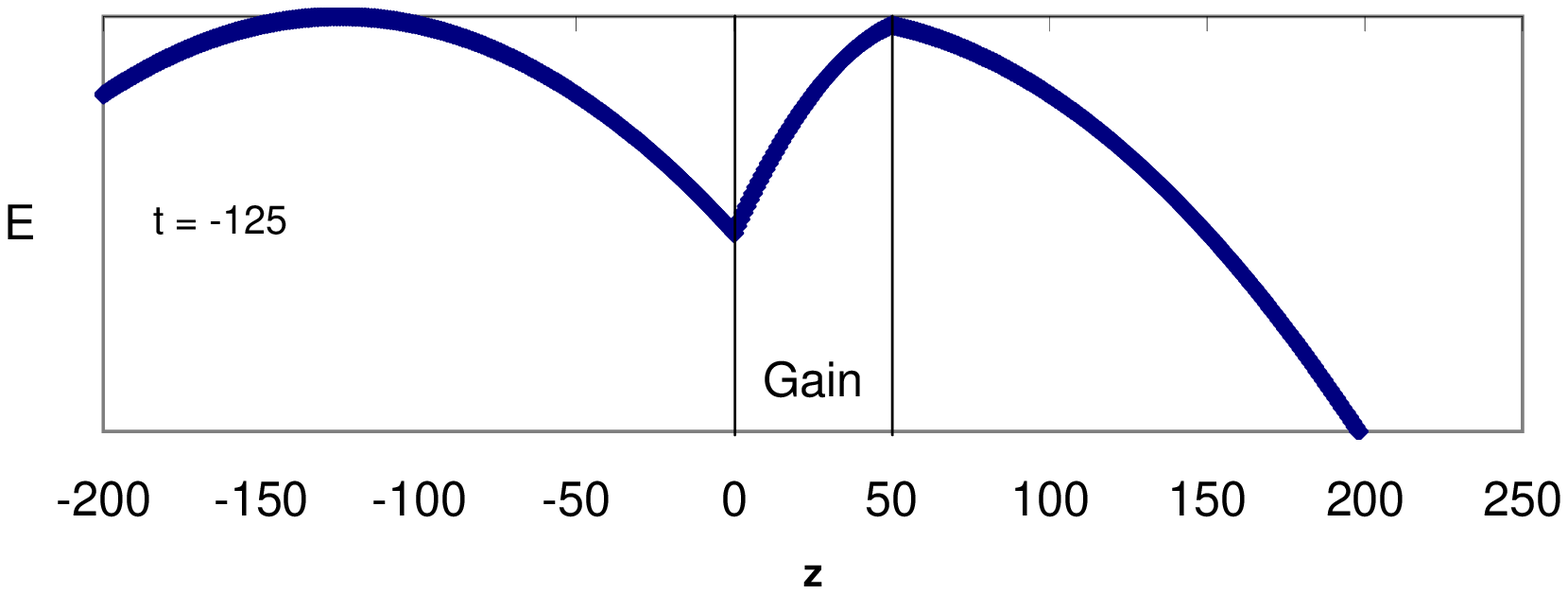} \\
\includegraphics*[width=2.75in, bb = 56 355 531 486]{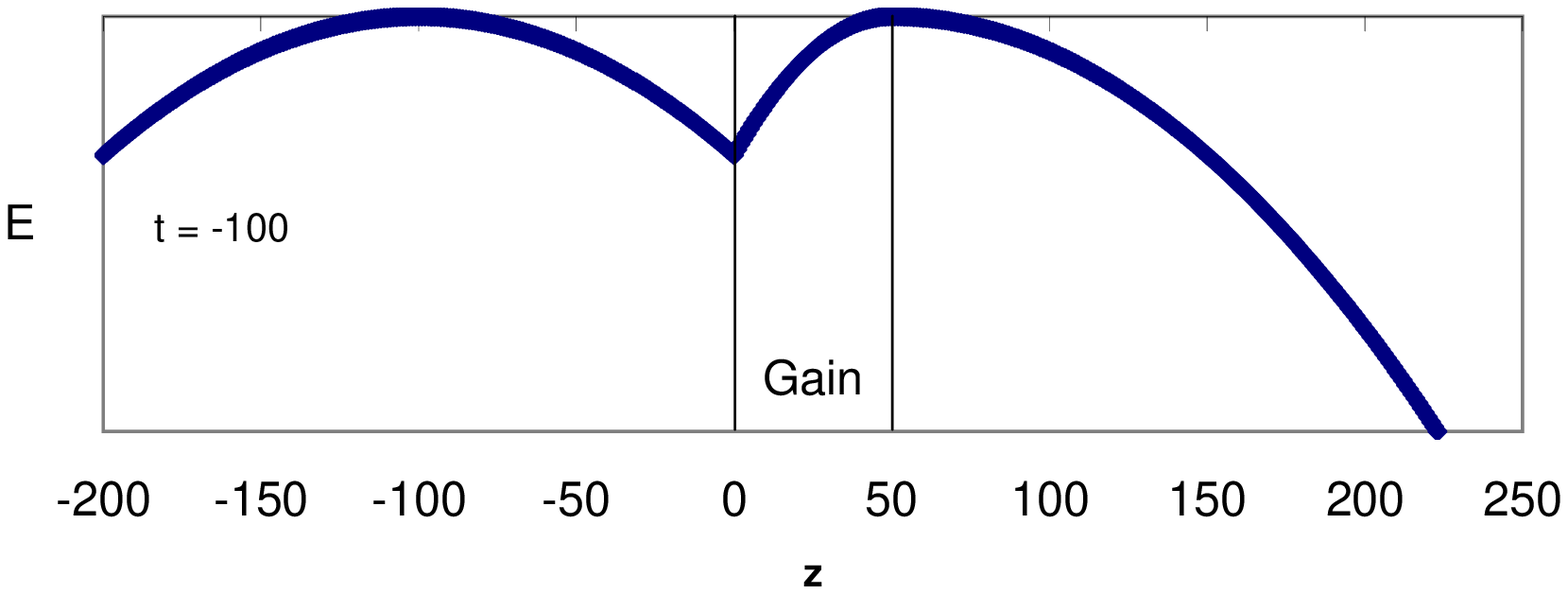} \\
\includegraphics*[width=2.75in, bb = 56 355 531 487]{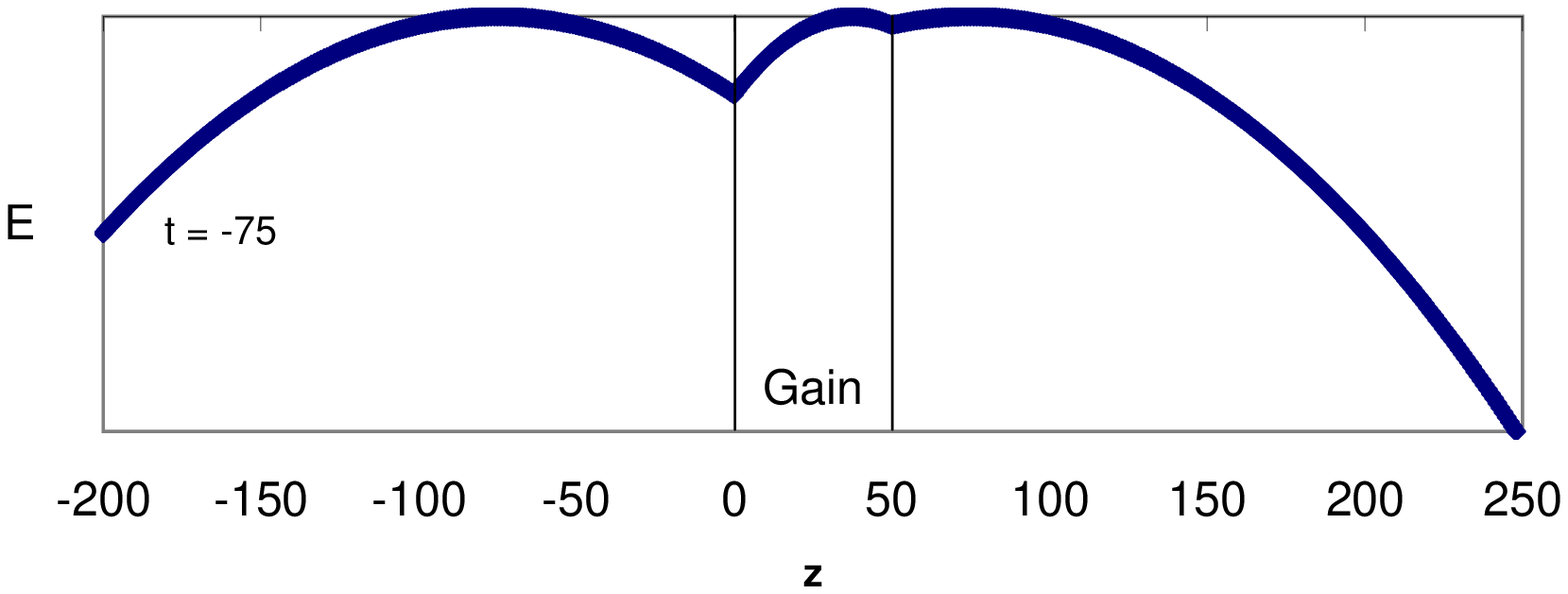} \\
\includegraphics*[width=2.75in, bb = 56 355 531 487]{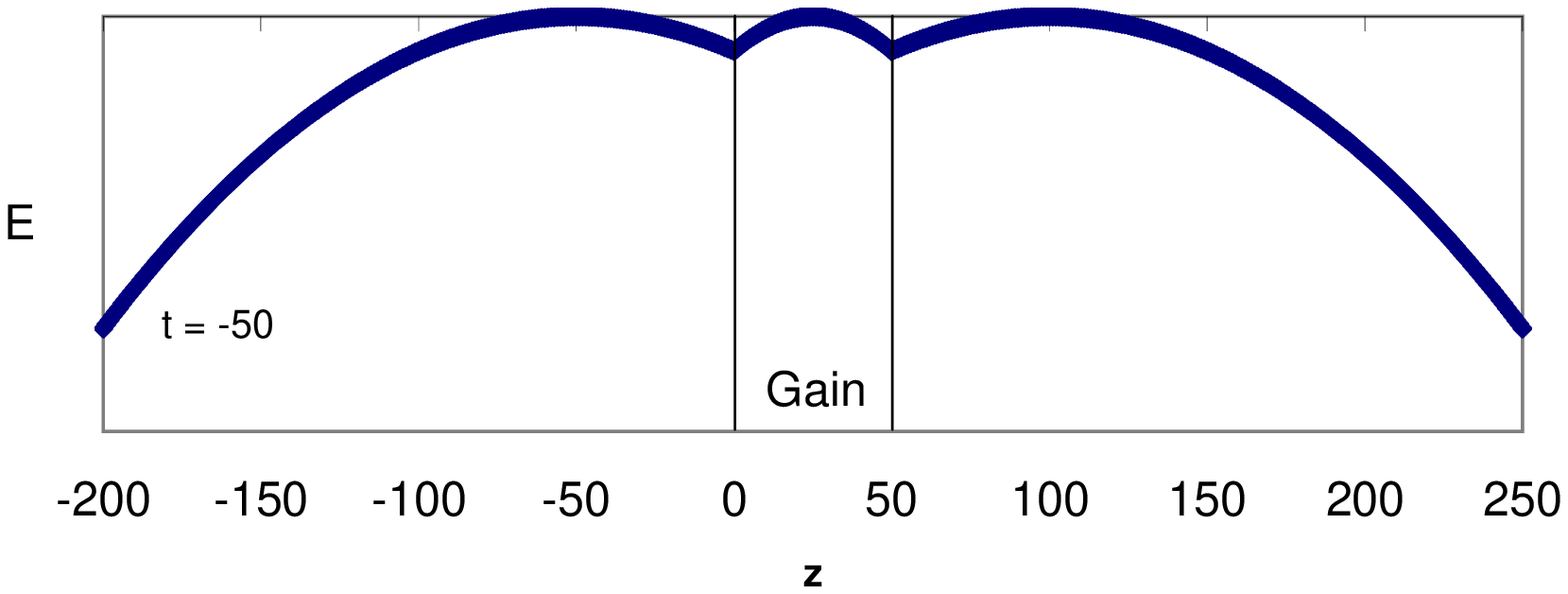} \\
\includegraphics*[width=2.75in, bb = 56 355 531 486]{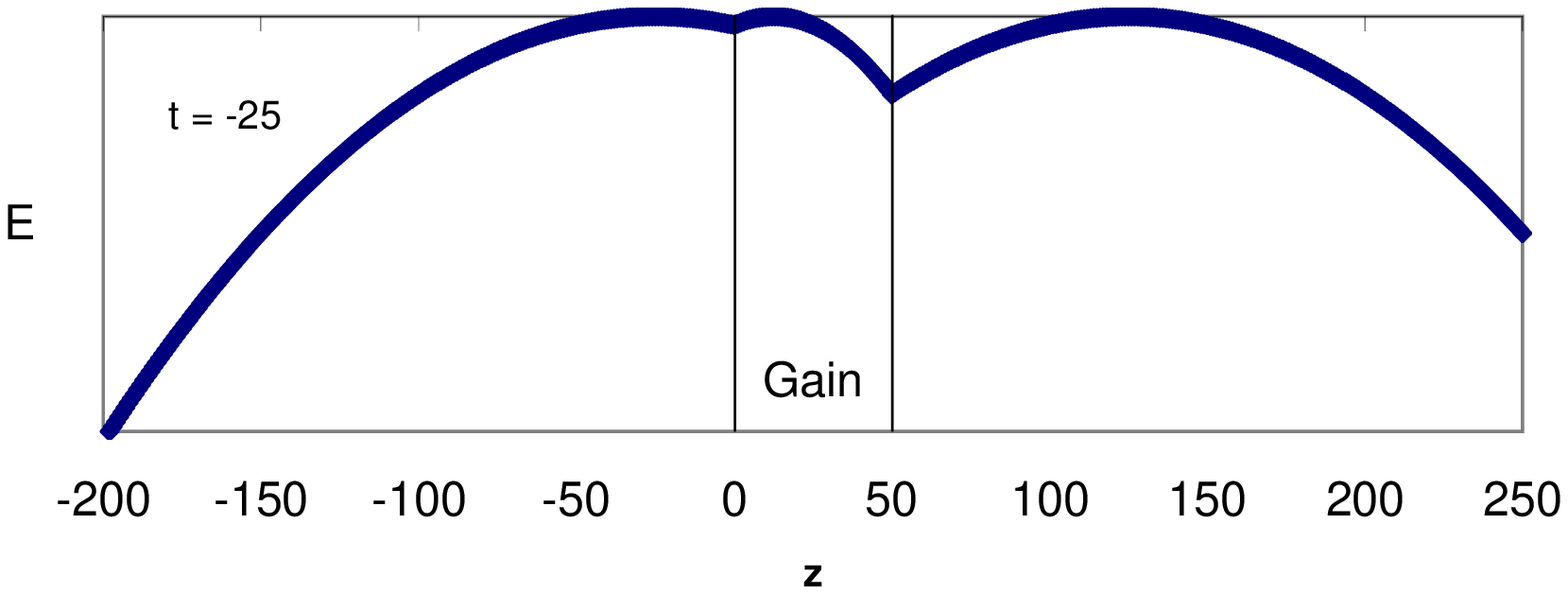} \\
\includegraphics*[width=2.75in, bb = 56 355 531 490]{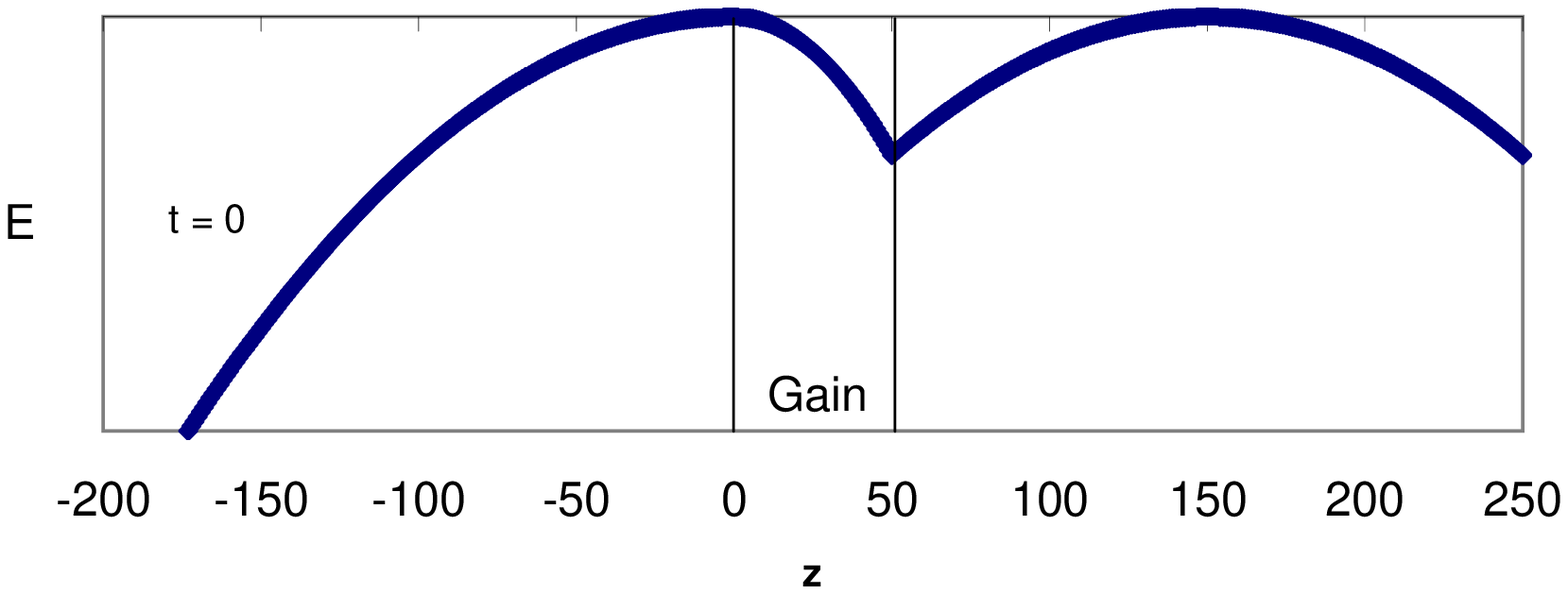} \\
\includegraphics*[width=2.75in, bb = 56 355 531 487]{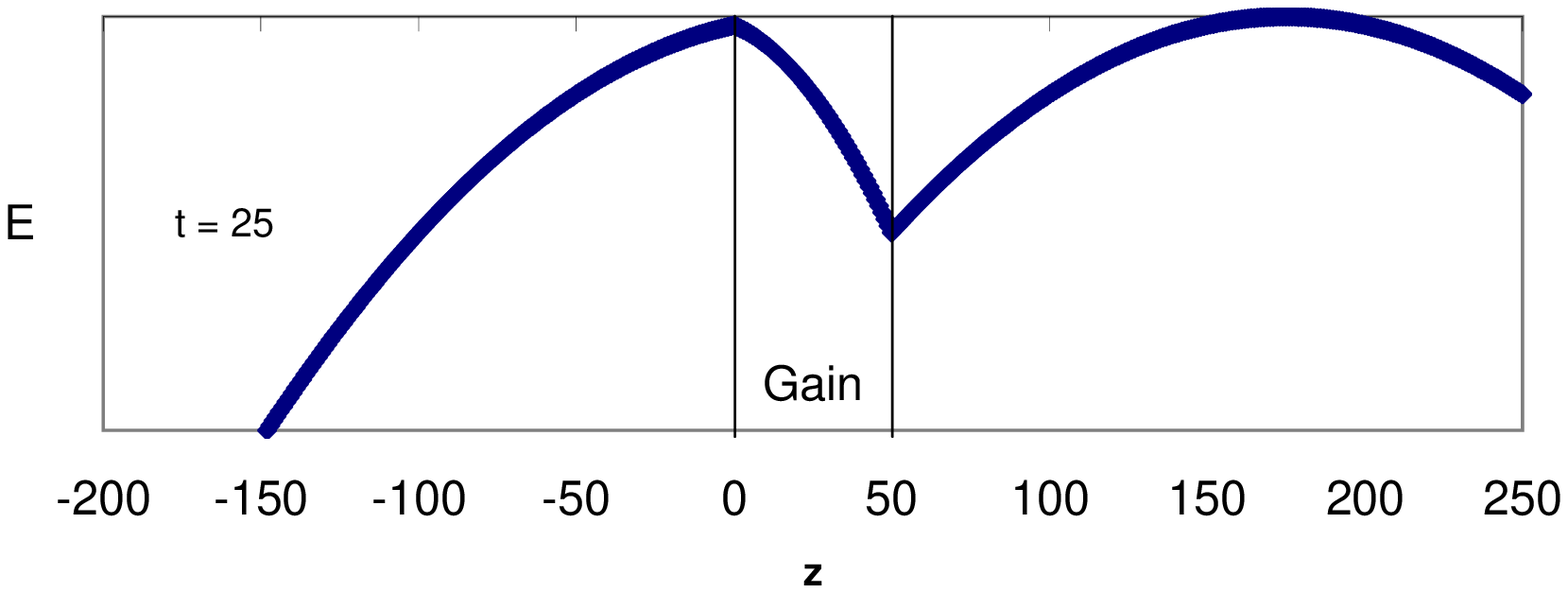} \\
\includegraphics*[width=2.75in, bb = 56 305 531 488]{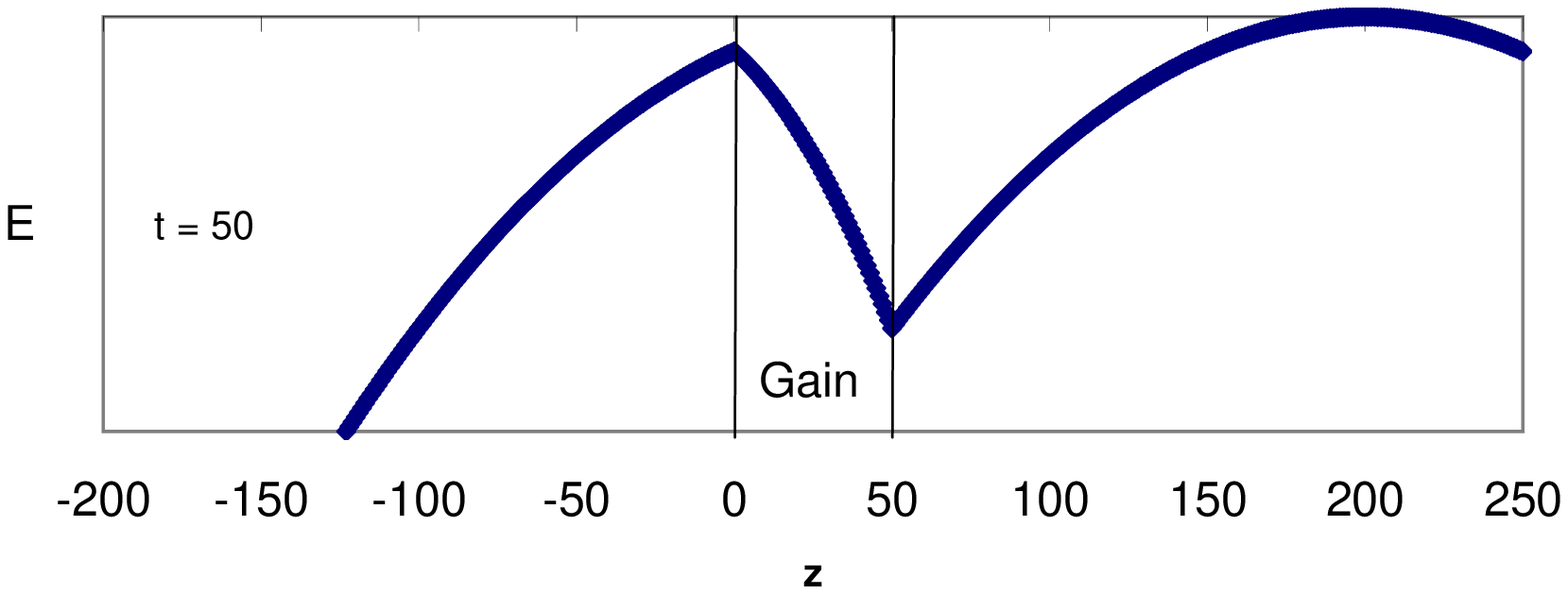}
\parbox{5.5in} 
{\caption[ Short caption for table of contents ]
{\label{fig3} The same as Fig.~\ref{fig2}, but with the electric field plotted
on a logarithmic scale from 1 to $10^{-65}$.
}}
\end{center}
\end{figure}

The effect of the negative group velocity medium can be dramatized in a
calculation based on eq.~(\ref{eq18c})
in which the pulse width is narrower than the gain region (in violation of
condition (\ref{eq26})), as 
shown in Fig.~\ref{fig2}.  Here, 
the gain region is
$0 < z < 50$, the group velocity is taken to be $-c/2$,
and $c$ is defined to be unity.
The behavior illustrated in Fig.~\ref{fig2} is perhaps less surprising
when the pulse amplitude is plotted on a logarithmic scale, as in 
Fig.~\ref{fig3}.  Although the overall gain of the system is near unity, the
leading edge of the pulse is amplified by about 70 orders of magnitude in
this example (the implausibility of which underscores that condition
(\ref{eq26}) cannot be evaded), while 
the trailing edge of the pulse is attenuated by the same amount.
The gain medium has temporarily loaned some of its energy to the pulse
permitting the leading edge of the pulse to appear to advance faster than
the speed of light.

Our discussion of the pulse has been based on a classical analysis of 
interference, but, as remarked by Dirac \cite{Dirac}, 
classical optical
interference describes the behavior of the wave functions of
individual photons, not of interference between photons.  Therefore, we
expect that the
behavior discussed above will soon be demonstrated for a ``pulse''
 consisting of a single photon with a Gaussian wave packet.  

\bigskip

The author thanks Lijun Wang for discussions of his experiment, and
Alex Granik for references to the early history
of negative group velocity and for the analysis contained in 
eqs.~(\ref{eq13})-(\ref{eq13c}).


\begin{thebibliography}{99}
\bibitem{note}
This is in contrast to the ``$\Lambda$'' configuration of the three-level
atomic system required for
slow light \cite{slowlight} where the pump laser does not produce an
inverted population, in which case an adequate classical description is
simply to reverse the sign of the damping constant for the pumped oscillator.
\end{thebibliography}
\end{document}